\documentclass[pra,twocolumn,showpacs,preprintnumbers,superscriptaddress]{revtex4}
\usepackage{amsmath}
\usepackage{amssymb}
\usepackage{graphicx}
\usepackage{epstopdf}
\usepackage{textcomp}
\usepackage{subfigure}
 \usepackage{mathrsfs}
\usepackage{color}
\newtheorem{theorem}{Theorem}[section]

\DeclareMathOperator{\tr}{Tr}
\DeclareMathOperator{\di}{diag}
\begin{document}

\title{Broadcasting of Quantum Correlations: Possibilities \& Impossibilities}
\author{Sourav Chatterjee}
\affiliation{Center for Computational Natural Sciences and Bioinformatics, International Institute of Information 
Technology-Hyderabad, Gachibowli, Telangana-500032, India.} 
\author{Sk Sazim}
\affiliation{Institute of Physics, Sainik School Post, Bhubaneswar-751005, Odisha, India}
\author{Indranil Chakrabarty}
\affiliation{Center for Security, Theory and Algorithmic Research, International Institute of Information 
Technology-Hyderabad, Gachibowli, Telangana-500032, India.}

\begin{abstract}

In this work, we extensively study the problem of broadcasting of quantum correlations (QCs). This includes broadcasting of quantum entanglement as well as correlations that go beyond the notion of entanglement (QCsbE). It is quite well known from the ``No-Broadcasting theorem" that perfect broadcasting of QCs is not possible. However it does not rule out the possibility of partial broadcasting of QCs where we can get lesser correlated states from a given correlated state. In order to have a holistic view of broadcasting, we investigate this problem by starting with most general representation of two qubit mixed states in terms of the Bloch vectors. As a cloning transformation we have used universal symmetric optimal Buzek-Hillery (B-H) cloner both locally and nonlocally. Unlike entanglement, we find that it is impossible to broadcast QCsbE optimally. 
Lastly, we generalize these results for any  symmetric or asymmetric cloning machines as well. This result brings out a fundamental difference between the correlations defined from the perspective of entanglement and the correlations measure which claims to go beyond entanglement.
\end{abstract}

\maketitle

\section{Introduction}

The impossibility  to clone quantum states is regarded as one of the most fundamental restriction which nature provides us \cite{wootters}. The ``No cloning theorem" states that there exists no quantum mechanical process that can take two different non-orthogonal quantum states $|\psi_1 \rangle$, $|\psi_2\rangle$ into states $|\psi_1\rangle\otimes|\psi_1\rangle$, $|\psi_2\rangle\otimes|\psi_2\rangle$ respectively. 
Even though we cannot copy an unknown quantum state perfectly but quantum mechanics never rules out the possibility of cloning it approximately \cite{wootters,buzek1,brub,buzek3,cloning-others,gisin,cerf1,cerf2,clone-review}. It also allows probabilistic cloning as one can always clone an arbitrary quantum state perfectly with some non-zero probability of succsess \cite{duan, clone-review}.

In the year 1996,  Buzek et al. introduced the concept of approximate cloning with certain fidelity. In this process, the state independent quantum copying machine was introduced by keeping the fidelity of cloning independent of the input state parameters. This machine is popularly known as universal quantum cloning machine (UQCM) \cite{buzek1} which was later proven to be optimal \cite{brub, gisin2}. Apart from this state independent quantum cloning machine (QCM), there are also state dependent QCMs for which the quality of copies depend on the input state \cite{brub, adhikari, clone-review}. 

Quantum entanglement \cite{einstein} which lies at the heart of quantum information theory is one of the key factor for better achievement of fidelity of QCMs \cite{sazim-cloning}. Not only that, it also plays a significant role in computational and communicational processes like quantum key distribution \cite{ekert, ben1}, secret sharing \cite{secretSharing}, teleportation \cite{ben3}, superdense coding \cite{ben2}, entanglement swapping \cite{horodecki-entanglement, bose}, remote entanglement distribution \cite{sazim-tele} and in many more tasks \cite{entanglement-others}. 
Atleast in the context of quantum  information processing, purer the entanglement, more valuable is the given two qubit state. Therefore, extraction of pure quantum entanglement from a partially entangled state is considered to be an important task. Consequently, there have been a lot of work on purification procedures by many researchers over the last few years showing how one can compress the amount of quantum entanglement locally \cite{ben4, compression}. The possibility of compression of quantum correlations naturally raises the question if the opposite i.e. decompression of correlations is realizable or not? Many researchers have answered this query using the process known as ``Broadcasting of Inseparability" \cite{buzek2, kar, adhikari}. This question becomes important when there is an exigency in increasing the number of available entangled pairs rather than the purity of it. In simple sense, broadcasting here refers to local or nonlocal copying of quantum correlations \cite{buzek2, satya-broad2}.

In general, the term broadcasting can be used in different contexts. Classical theory permits broadcasting of  information, however that is not the case for all states in quantum theory. Cloning and broadcasting principles demarcate the boundary between classical and quantum worlds. In this context, Barnum \textit{et al} were the first to show that non-commuting mixed states do not meet the criteria of broadcasting \cite{barnum-noncommu}.

It is impossible to have a process which will perfectly copy (clone and broadcast) an arbitrary quanutm state \cite{wootters, buzek2, barnum-noncommu}.  By referring to perfect broadcasting of correlations we mean that the correlations in a two qubit state $\rho^{ab}$ are locally broadcastable if there exist two operations, $\Sigma^a$: $S(\mathbb{H}^a) \rightarrow S(\mathbb{H}^{a_{1}}\otimes \mathbb{H}^{a_{2}})$ and $\Sigma^b$: $S(\mathbb{H}^b) \rightarrow S(\mathbb{H}^{b_{1}} \otimes \mathbb{H}^{b_{2}})$ such that $I(\rho^{a_1b_1})$ = $I(\rho^{a_2b_2})$ = $I(\rho^{ab})$. Here, $I(\rho^{ab})$ 
is the quantum mutual information, $\rho^{a_1a_2b_1b_2}:=\Sigma^a \otimes \Sigma^b (\rho^{ab})$ and $\rho^{a_ib_i}:=\tr_{a_{\bar{i}}b_{\bar{i}}}(\rho^{a_1a_2b_1b_2})$ 
\cite{piani}. Quite recently, many authors showed that correlations in a single bipartite state can be locally or unilocally broadcast if and only if the states are classical (i.e. having classical correlations) or classical-quantum respectively \cite{piani, barnum-gen, luo, luo-li}.

In the previous cases, we generally discussed about broadcasting of a general quantum state or perfect broadcasting of correlations. But when we refer broadcasting of an entangled state, we generally talk about creating more pairs of lesser entangled states from a given entangled state where $I(\rho^{a_1b_1})$ and $I(\rho^{a_2b_2})$ are less than $I(\rho^{ab})$. This is done via the application of local cloning operation on each qubit of the given entangled state, or sometimes by applying global cloning operations on the total input entangled state itself \cite{buzek3, buzek2, kar}. Bandyopadhyay \textit{et al.} \cite{kar} showed that only UQCMs having fidelity over $\frac{1}{2}(1+\frac{1}{\sqrt{3}})$ can broadcast entanglement and further that entanglement in the input state is optimally broadcast only if the quantum cloners used for local copying  are optimal. However, the fact that if local cloners are used then broadcasting of entanglement into more than two entangled pairs is impossible. Ghiu \textit{et al.}
addressed the question of  broadcasting of entanglement by using local universal optimal asymmetric Pauli cloning machines. They presented that if one employs symmetric cloners instead of asymmetric ones, then only optimal broadcasting of inseparability is achievable \cite{ghiu}. In other works, authors investigated the problem of secretly broadcasting of three-qubit entangled state between two distant partners with universal quantum cloning machine and then the result is generalized to generate secret entanglement among three parties \cite{satya-broad2}. Various other works on broadcasting of entanglement depending on the types of QCMs were also done in the later period \cite{indranil-broad, satya-continuous-broad}. 

In this work, we mainly investigate the problem of broadcasting of quantum correlations (QCs). Traditionally, by QCs we refer to entanglement. First part of our study is about broadcasting of quantum entanglement for general two qubit mixed states. For the first time in the existing research on broadcasting, we provide the broadcasting range for general two qubit state in terms of Bloch vectors. To do this we apply the Buzek-Hillery (B-H) QCM, both locally and non-locally. We separately provide broadcasting ranges for werner-like and Bell-diagonal states as illustration. In the second part of our work, while exploring the possibility of broadcasting of quantum correlations that go beyond entanglement (QCsbE), remarkably we find that it is impossible to broadcast optimally such correlations with the help of any local or nonlocal cloners. We analytically prove this by first taking the B-H state dependent and independent cloners and then by logically extending our result for the other cloners as well. This is indeed 
one such result which highlights how fundamentally two approaches, QCsbE and entanglement, are different. However, we can broadcast QCsbE if we relax the optimality conditions.

In section II, we first introduce the quantum cloning machines, more specifically the state independent and 
dependent versions of B-H cloners, which we will later use for our local as well as nonlocal cloning processes. 
In section III, we define broadcasting of entanglement via local cloning operations as well as non-local cloning operation and then obtain the generalized optimal broadcasting range for any two qubit state in terms of Bloch vectors. 
In each of the two above cases, we exemplify our results for two types of mixed states: namely the Werner-like and the  Bell-diagonal states. 
In section IV, we give the definition for broadcasting of QCsbE and explicitly discuss the possibilities and impossibilities of such broadcasting.
Lastly, in section V, we conclude with a small conjecture by which broadcasting of correlations beyond entanglement might be possible.
\section{Quantum cloning machines beyond No-cloning theorem}
\label{sec:cloningmachines}
Quantum cloning transformations can be viewed as a completely positive (CP) trace preserving map between two quantum systems, supported by an ancilla \cite{brub, clone-review}. In this section, we briefly describe the Buzek-Hillery (B-H) QCM which we will later use for analysing the possibility and impossibility of broadcasting of entanglement as well as correlations beyond entanglement respectively. 

B-H cloning machine ($U_{bh}$) is a $M$-dimensional quantum copying transformation acting on a state $\left|\Psi_i\right\rangle_{a_0}$ ($i$ = 1, ..., $M$). This state is to be copied on a blank state $\left|0\right\rangle_{a_1}$. The copier is initially prepared in state $\left|X\right\rangle_x$ which subsequently get transformed into another set of state vectors $\left|X_{ii}\right\rangle_x$ and $\left|Y_{ij}\right\rangle_x$ as a result of application of the cloner. Here $a_0$, $a_1$ and $x$ represent the input, blank and machine qubits respectively. In this case, these transformed state vectors belong to the orthonormal basis set in the $M$-dimensional space. The transformation scheme $U_{bh}$ is given by \cite{buzek3},
\begin{eqnarray}
&&U_{bh}\left|\Psi_i\right\rangle_{a_0} \left|0\right\rangle_{a_1} \left|X\right\rangle_x \rightarrow  c\left|\Psi_i\right\rangle_{a_0} \left|\Psi_i\right\rangle_{a_1} \left|X_{ii}\right\rangle_x \nonumber\\
&&+d\displaystyle \sum_{j\neq i}^{M} \left(\left|\Psi_i\right\rangle_{a_0} \left|\Psi_j\right\rangle_{a_1} +\left|\Psi_j\right\rangle_{a_0}\left|\Psi_i\right\rangle_{a_1}\right) \left|Y_{ij}\right\rangle_x,
\label{eq:B-H_gen_transform}
\end{eqnarray}
where $i,\:j$ = $\{1,...,M\}$,  and the coefficients $c$ and $d$ are real. 
\subsection{State independent cloning transformations}
An optimal state independent version of the B-H cloner ($U_{bhsi}$) can be obtained from Eq.\eqref{eq:B-H_gen_transform} by imposing the unitarity and normalization conditions which give rise to the following constraints,
\begin{eqnarray}
& \left\langle X_{ii}|X_{ii}\right\rangle=\left\langle Y_{ij}|Y_{ij}\right\rangle=\left\langle X_{ii}|Y_{ji} \right\rangle=1,
\label{eq:coefficients}
\end{eqnarray}
when $\left\langle X_{ii}|Y_{ij} \right\rangle=\left\langle Y_{ji}|Y_{ij} \right\rangle=\left\langle X_{ii}|X_{jj}\right\rangle=0$, with $i\neq j$ and $c^2=\frac{2}{M+1}$, $d^2 =\frac{1}{2(M+1)}$. Here, we consider $M=2^m$ where $m$ is the number of qubits in a given quantum register. In the above transformation, by demanding the independence of the scaling (shrinking) property on input state parameters it is ensured that the quality of the cloning (fidelity of the output copies) doesn't depend on the input state \cite{brub, buzek3}. 
\subsubsection{Local state independent cloner}
The above optimal cloner $U_{bhsi}$ with $M=2$ becomes a local copier ($U^{l}_{bhsi}$). From Eq.~\eqref{eq:coefficients} it can be easily observed that the corresponding values of coefficients $c$ and $d$ become $\sqrt{\frac{2}{3}}$ and $\sqrt{\frac{1}{6}}$ respectively. By substituting these values of the coefficients in Eq.~\eqref{eq:B-H_gen_transform}, we can obtain the optimal state independent cloner which can be used for local copying purposes\cite{buzek2}.
\subsubsection{Nonlocal state independent cloner}
When $M=4$ the above optimal cloner $U_{bhsi}$ becomes a nonlocal copier ($U^{nl}_{bhsi}$). Then the corresponding values of the coefficients $c$ and $d$ in Eq.~\eqref{eq:coefficients} become $\sqrt{\frac{2}{5}}$ and $\sqrt{\frac{1}{10}}$ respectively. By substituting these coefficients in $U_{bh}$ given by Eq.~\eqref{eq:B-H_gen_transform}, we can obtain the optimal state independent cloner used for nonlocal copying purposes \cite{buzek3}.
\subsection{State dependent cloning transformations}
The B-H state dependent cloner ($U_{bhsd}$) was developed from this B-H state independent cloning transformation ($U_{bhsi}$), given in Eq.~\eqref{eq:B-H_gen_transform} with $U_{bh}=U_{bhsi}$, by relaxing the universality condition: $\frac{\partial D}{\partial<inp>} = 0$; where $<inp>$ represents all the parameters of the input state. The distortion $D$ describes the distance between the input and output states of the cloner \cite{adhikari}.

With $c$ = $d$ = 1, the unitarity constraints on the B-H cloning transformation in Eq.~\eqref{eq:B-H_gen_transform} give rise to the following conditions on the output states, which are no longer necessarily orthonormal, 
\begin{eqnarray}
 & \left\langle X_{ii}|X_{ii} \right\rangle + \displaystyle \sum^{M}_{j\neq i} 2 \left\langle Y_{ij}|Y_{ij} \right\rangle =1,\:\left\langle Y_{ij}|Y_{kl} \right\rangle =0
\end{eqnarray}
where $i \neq j$ and $ij \neq kl$ for  $i,j,k,l=\{1,...,M\}$. We assume that,
$\left\langle X_{ii}|Y_{jk} \right\rangle = \frac{\mu}{2}$, $\left\langle Y_{ij}|Y_{ij} \right\rangle=\lambda$, $\left\langle X_{ii}|X_{jj} \right\rangle=\left\langle X_{ii}|Y_{ij} \right\rangle = 0$,
where again $i \neq j$ for  $i,j,k=\{1,...,M\}$; $\mu$ and $\lambda$ are the machine parameters. 
By equating the dependence of the distortion $D$ on the machine parameter $\lambda$ to zero, in each of the cases, we can calculate the value of $\lambda$ for which the B-H state dependent cloners become optimal with respect to that ensemble of input states.
\subsubsection{Local state dependent cloner}
For the case of a local state dependent cloner ($U^{l}_{bhsd}$), the distortion $D$ is $D_{ab}$ = $\tr[ \rho_{ab}^{(out)}-\rho_{a}^{(id)}\otimes \rho_{b}^{(id)}]^2$. If $|\psi_{a(b)}^{(id)}\rangle=\alpha|0\rangle_{a(b)}+\beta|1\rangle_{a(b)}$ be an arbitrary pure state of one 
qubit in mode ``$a$''  or ``$b$'', where $\alpha, \beta$ represents the input state parameters with $\alpha^2+\beta^2=1$ being the normalization condition; then $\rho_{a}^{(id)}=|\psi_a^{(id)}\rangle\langle\psi_a^{(id)}|$ and $\rho_{b}^{(id)}=|\psi_b^{(id)}\rangle\langle\psi_b^{(id)}|$ represents output modes in case of an ideal copy. However, in a more realistic situtation when cloning fidelity is non-ideal then the output state of the cloner is given by $\rho_{ab}^{(out)}$. Solving the equation $\frac{\partial D_{a}}{\partial\alpha^{2}} = 0$, where $D_{a}$ = $\tr[ \rho_{a}^{(out)}-\rho_{a}^{(id)}]^2$; with $\rho_{a}^{(out)}=\tr_{b}[\rho_{ab}^{(out)}]$, we can derive the relation between the parameters $\lambda$ and $\mu$. It turns out to be $\mu = 1-2\lambda$. So the permitted range of $\lambda$ is bounded by $\{0, \frac{1}{2}\}$ in this case. However, it can be noted that here the value $\lambda=\frac{1}{6}$ is restricted, since for such values it reduces to the B-H optimal state independent local cloner 
$U^l_{bhsi}$ and consequently looses the input state dependence property. 
\subsubsection{Nonlocal state dependent cloner}
For the case of a nonlocal state dependent cloner ($U^{nl}_{bhsd}$), the distortion $D$ is 
$D_{abcd}$ = $\tr[ \rho_{abcd}^{(out)}-\rho_{ab}^{(id)}\otimes \rho_{cd}^{(id)}]^2$. 
If $|\phi_{ab(cd)}^{(id)}\rangle=\alpha|00\rangle_{ab(cd)}+\beta|11\rangle_{ab(cd)}$ 
be the non-maximally entangled state of two qubits in mode ``$ab$''  or ``$cd$''; then $\rho_{ab}^{(id)}=|\psi_{ab}^{(id)}\rangle\langle\psi_{ab}^{(id)}|$ and 
$\rho_{cd}^{(id)}=|\psi_{cd}^{(id)}\rangle\langle\psi_{cd}^{(id)}|$ represents output modes in case of an ideal copy. However, in a more realistic situtation when cloning fidelity is non-ideal then the output state of the cloner is given by $\rho_{abcd}^{(out)}$. Solving the equation $\frac{\partial D_{ab}}{\partial\alpha^{2}} = 0$, where $D_{ab}$ = $\tr[ \rho_{ab}^{(out)}-\rho_{ab}^{(id)}]^2$; with $\rho_{ab}^{(out)}=\tr_{c,d}[\rho_{abcd}^{(out)}]$, we can derive the relation between the parameters $\lambda$ and $\mu$. 
Here, it turns out to be $\mu = 1-4\lambda$. So the permitted range of $\lambda$ is bounded by $\{0, \frac{1}{4}\}$ in this case. 
However, it can be noted that the value $\lambda=\frac{1}{10}$ is restricted, since for such values it reduces to the B-H optimal 
state independent nonlocal cloner $U^{nl}_{bhsi}$ thereby loosing the input state dependence property.
\section{Broadcasting of Quantum Entanglement}
\label{sec:broad_entanglement}
In this section, we consider broadcasting of quantum entanglement (inseparability) with the help of both local and nonlocal cloning operations. Let us begin with a situation where we have two distant parties $A$ and $B$ and they share a two qubit mixed state $\rho_{12}$ which can be canonically expressed as \cite{gisin2}:
\begin{eqnarray}
\rho_{12}&=&\frac{1}{4}[\mathbb{I}_4+\sum_{i=1}^{3}(x_{i}\sigma_{i}\otimes \mathbb{I}_2+ y_{i}\mathbb{I}_2\otimes\sigma_{i})\nonumber\\
&+&\sum_{i,j=1}^{3}t_{ij}\sigma_{i}\otimes\sigma_{j}]=\left\{\vec{x},\:\vec{y},\: T\right\}\:\:\: \mbox{(say),} \label{eq:mix}
\end{eqnarray}
where $x_i=\tr[\rho_{12}(\sigma_{i}\otimes \mathbb{I}_2)]$,  $y_i=\tr[\rho_{12}(\mathbb{I}_2\otimes\sigma_{i})]$ and $t_{ij}=\tr[\rho_{12}(\sigma_i\otimes\sigma_{j})]$ with [$\sigma_i;\:i$ = $\{1,2,3\}$] are $2\otimes 2$ Pauli matrices and $\mathbb{I}_n$ is the identity matrix of order $n$. And $\vec{x}=\left\{ x_{1},\: x_{2},\: x_{3}\right\}$, $\vec{y}=\left\{ y_{1},\: y_{2},\: y_{3}\right\}$ are Bloch coloumn vectors and $T=[t_{ij}]$ is the correlation matrix.
 
In order to test the separability as well as inseparability for the bipartite states, we generally use Peres-Horodecki criteria. This is a necessary and sufficient condition for detection of entanglement for bipartite systems with dimension $2 \otimes 2$ and $2 \otimes 3$.

\noindent \emph{Peres-Horodecki criteria} \cite{peres}: 
If atleast one of the eigenvalues of a partially transposed density operator for a bipartite state $\rho$ defined as $\rho_{m\mu,n\nu}^{T}=\rho_{m\nu,n\mu}$ turn out to be negative then we can say that the state $\rho$ is inseparable. 
This criteria can be equivalently expressed by the condition that at least one of the two determinants
\begin{eqnarray}
W_3=\begin{vmatrix}\rho_{00,00} & \rho_{01,00} & \rho_{00,10}\\
\rho_{00,01} & \rho_{01,01} & \rho_{00,11}\\
\rho_{10,00} & \rho_{11,00} & \rho_{10,10}
\end{vmatrix}\:\:\text{and}\nonumber\\
W_4=\begin{vmatrix}\rho_{00,00} & \rho_{01,00} & \rho_{00,10} & \rho_{01,10} \\
  \rho_{00,01} & \rho_{01,01} & \rho_{00,11} & \rho_{01,11} \\
  \rho_{10,00} & \rho_{11,00} & \rho_{10,10} & \rho_{11,10} \\
  \rho_{10,01} & \rho_{11,01} & \rho_{10,11} & \rho_{11,11} 
\end{vmatrix}
\label{eq:w3-w4}
\end{eqnarray}
is negative; with
$W_2=\begin{vmatrix}\rho_{00,00} & \rho_{01,00} \\
\rho_{00,01} & \rho_{01,01} \\
\end{vmatrix}$
being simultaneously non-negative.
\subsection{Broadcasting of entanglement via local and nonlocal cloning operations}
\begin{figure}[h]
\begin{center}
\[
\begin{array}{cc}
\includegraphics[height=4.5cm,width=4.5cm]{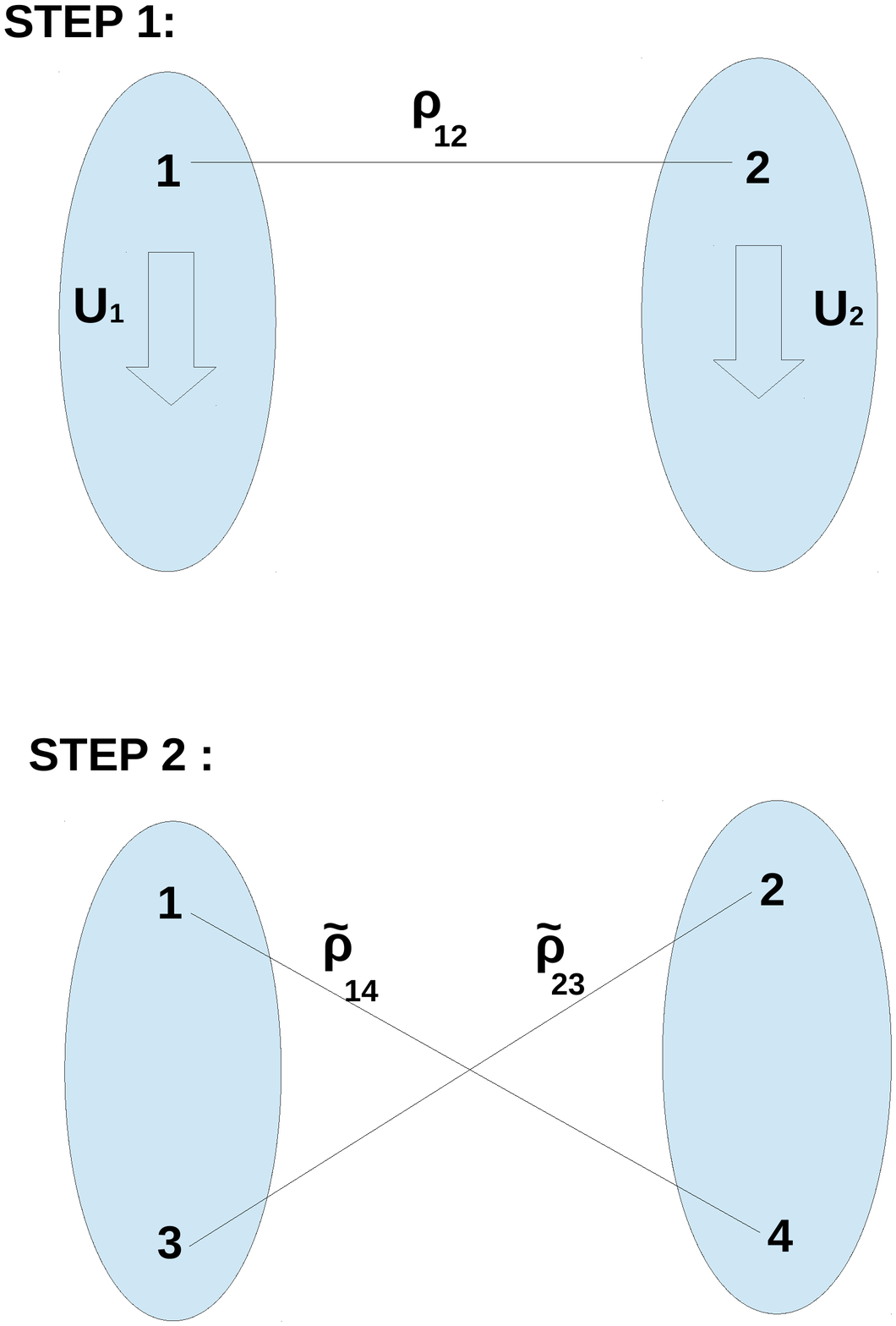}
\end{array}
\]
\end{center}
\caption{\noindent 
The figure shows the broadcasting of the state $\rho_{12}$ into $\tilde{\rho}_{14}$ and $\tilde{\rho}_{23}$ 
through application of local cloning unitaries $U_1$ and $U_2$ on both sides.\label{fig:local_ent}
}
\end{figure}

\noindent{\it Local cloning}: Each of the parties now individually apply a local copying operation on their own qubit i.e., $U_1 \otimes U_2$ to produce the state $\tilde{\rho}_{1234}$. 
The B-H state independent symmetric optimal cloning transformation ($U^l_{bhsi}$) used for local copying is obtained by putting $M=2$ in Eq.~\eqref{eq:B-H_gen_transform} with $c=\sqrt{\frac{2}{3}}$ and $d=\sqrt{\frac{1}{6}}$. The corresponding basis vectors are  $\left|\Psi_{1}\right\rangle =\left|0\right\rangle $ and $\left|\Psi_{2}\right\rangle =\left|1\right\rangle $. 
After we obtain the composite system $\tilde{\rho}_{1234}$, we trace out the qubits $2$, $4$ and $1$, $3$ to obtain the local output states $\tilde{\rho}_{13}(= \tr_{24}[U_1\otimes U_2(\rho_{12})])$ on A's side and $\tilde{\rho}_{24}(= \tr_{13}[U_1\otimes U_2(\rho_{12})])$ on B's side respectively. Similarly, after tracing out the local output states from the composite system, we have the nonlocal output states $\tilde{\rho}_{14}(=\tr_{23}[U_1\otimes U_2(\rho_{12})])$ and $\tilde{\rho}_{23}(=\tr_{14}[U_1\otimes U_2 (\rho_{12})])$ [see FIG. (\ref{fig:local_ent})]. 

\begin{figure}[h]
\begin{center}
\[
\begin{array}{cc}
\includegraphics[height=4.5cm,width=4.5cm]{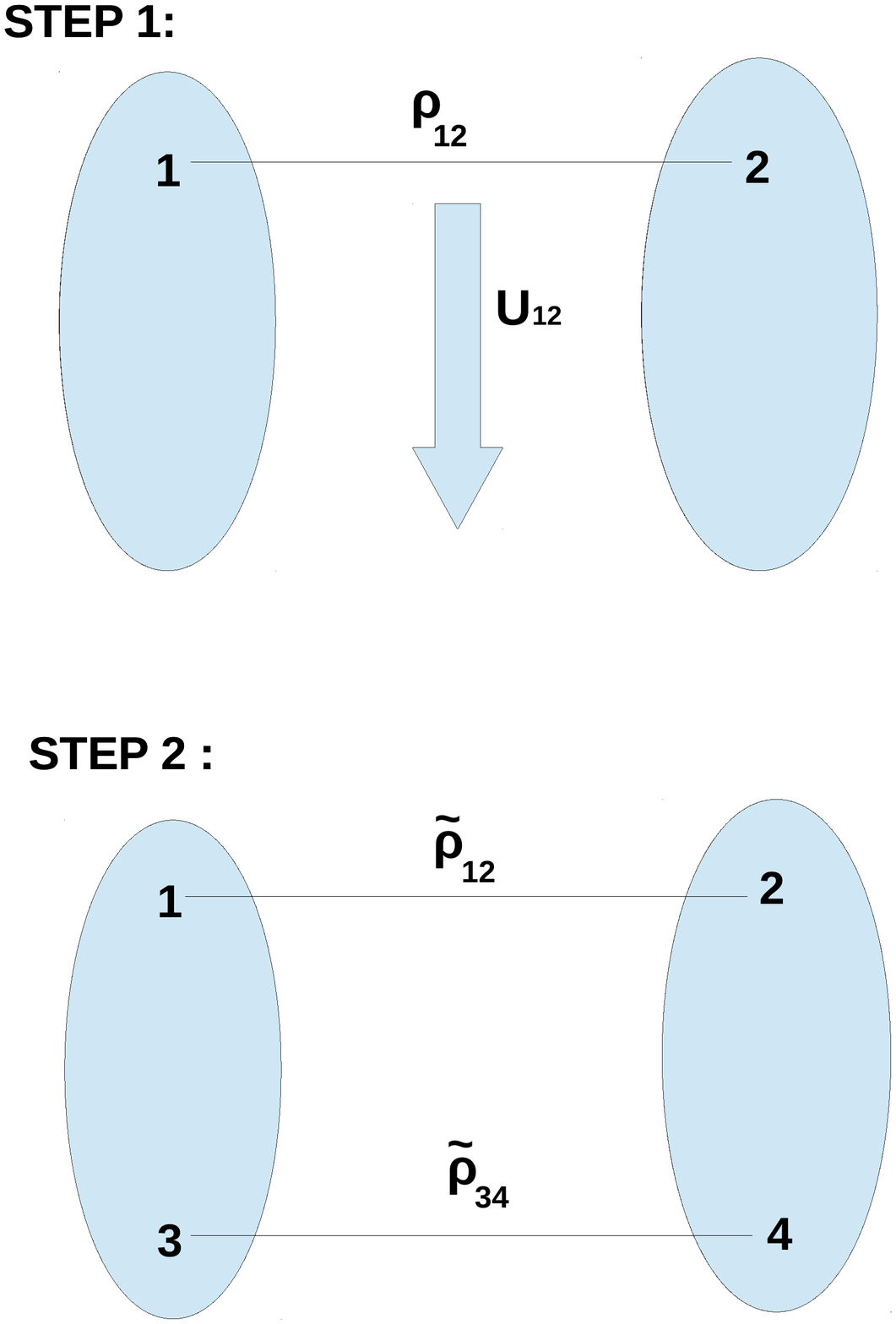}
\end{array}
\]
\end{center}
\caption{\noindent 
The figure shows the broadcasting of the state $\rho_{12}$ into $\tilde{\rho}_{12}$ and $\tilde{\rho}_{34}$ through application of a nonlocal (global) cloning unitary $U_{12}$. \label{fig:nonlocal}} 
\end{figure}

\noindent{\it Non-local cloning}: Here, the basic idea is that the entire state $\rho_{12}$ (given in Eq.~\eqref{eq:mix}) is in the same lab and the intension is to have more than one copy of it. In that process, we apply a global unitary operation $U_{12}$ to produce $\tilde{\rho}_{1234}$. The B-H state independent optimal cloning transformation ($U^{nl}_{bhsi}$) used for nonlocal copying is obtained by substituting $M=4$ in Eq.~\eqref{eq:B-H_gen_transform} with $c=\sqrt{\frac{2}{5}}$ and $d=\sqrt{\frac{1}{10}}$.
In this case, the corresponding basis vectors are  $\left|\Psi_{1}\right\rangle =\left|00\right\rangle $, $\left|\Psi_{2}\right\rangle =\left|01\right\rangle $, $\left|\Psi_{3}\right\rangle =\left|10\right\rangle $ and $\left|\Psi_{4}\right\rangle =\left|11\right\rangle $.   
%
Once we have the composite system $\tilde{\rho}_{1234}$, we trace out the qubits $3$ and $4$ to obtain the output state $\tilde{\rho}_{12}(=\tr_{34}[U_{12}\rho_{12}])$ or the qubits $1$ and $2$ to obtain $\tilde{\rho}_{34}(= \tr_{12}[U_{12}\rho_{12}])$. Next, proceeding in similar manner, we obtain the remaining states $\tilde{\rho}_{13}(= \tr_{24}[U_{12}\rho_{12}])$ and $\tilde{\rho}_{24}(= \tr_{13}[U_{12}\rho_{12}])$ by tracing out the qubits $2$, $4$ and $1$, $3$ from $\tilde{\rho}_{1234}$ respectively. We could have also chosen the diagonal pairs ($\tilde{\rho}_{14}\: \&\: \tilde{\rho}_{23}$) instead of choosing the pairs: $\tilde{\rho}_{12}\: \&\: \tilde{\rho}_{34}$ as our desired pairs. However, we refrain ourselves from choosing the pairs $\tilde{\rho}_{13}\: \&\: \tilde{\rho}_{24}$ as the desired pairs \cite{buzek3} [see FIG. (\ref{fig:nonlocal})]. 

In principle, to broadcast the amount of entanglement between the desired pairs $(1,4)/(1,2)$ and $(2,3)/(1,4)$ we just maximize the entanglement
 between the output pairs, regardless of the states between $(1,3)$ and
 $(2,4)$. However, for optimal broadcasting of entanglement across parties we require to minimize the amount of entanglement 
 within parties. 
This is because the total amount of entanglement ($E$) produced is the sum of the entanglement within parties ($E_{l}$) and the entanglement across 
the parties ($E_{nl}$),  i.e $E=E_{l}+E_{nl}$. The amount of entanglement ($E$) is strictly less or equal to the total entanglement of the 
input state. To maximize $E_{nl}$, we must have $E_{l}=0$.  In other words, for optimal 
 broadcasting we should have no entanglement between the qubits  $(1,3)$ and $(2,4)$.

\noindent\textbf{Definition 2.1:} 
An entangled state $\rho_{12}$ is said to be broadcast after the application of local cloning operation 
($U_1 \otimes U_2$), if for some values of the input state parameters, 
the states \{$\tilde{\rho}_{14}$, $\tilde{\rho}_{23}$\} are inseparable. 

\noindent \textbf{Definition 2.2:} An entangled state $\rho_{12}$ is said to be broadcast after the application of 
nonlocal cloning operation ($U_{12}$), if for some values of the input state parameters, the desired output 
states \{$\tilde{\rho}_{12}$, $\tilde{\rho}_{34}$\} are entangled.

\noindent \textbf{Definition 2.3:}  
An entangled state $\rho_{12}$ is said to be broadcast optimally after the application of local cloning operation ($U_1 \otimes U_2$), 
if for some values of the input state parameters, the  states  \{$\tilde{\rho}_{14}$, $\tilde{\rho}_{23}$\} are inseparable 
and the states \{$\tilde{\rho}_{13}$, $\tilde{\rho}_{24}$\} are separable.

\noindent\textbf{Definition 2.4:}  
An entangled state $\rho_{12}$ is said to be broadcast optimally after the application of nonlocal cloning operation 
($U_{12}$), if for some values of the input state parameters, the desired output 
states \{$\tilde{\rho}_{12}$, $\tilde{\rho}_{34}$\} are entangled, 
and the remaining output states \{$\tilde{\rho}_{13} (= \tr_{24}[U_{12}\rho_{12}])$, $\tilde{\rho}_{24}(= \tr_{13}[U_{12}\rho_{12}])$\} are separable.

If we consider the non-optimal broadcasting then the broadcasting range will increase 
whereas for optimal one the broadcasting range will be small. Let us consider a general pure two-qubit state in 
Schmidt form $|\psi_{12}\rangle=\sqrt{\lambda}|00\rangle\langle 00|+\sqrt{1-\lambda}|11\rangle\langle11|$, where $\lambda$ is Schmidt coefficient and $0\leq\lambda\leq 1$. Now if we apply B-H local cloning operation ($U_1 \otimes U_2$) on this state, the local output states will only be separable when $L_-<\lambda<L_+$, where $L_{\pm}=\frac{1}{16}(8\pm\sqrt{39})$ \cite{buzek2} and hence it is the optimal broadcasting range. If we relax the optimality condition i.e., $E_l\neq 0$
then we can easily conclude that the broadcasting of entanglement may be possible for greater range 
of $\lambda$. 
The same analysis is applicable for non-local cloning  and same type of feature will appear. 
Next, we will discuss the optimal broadcasting of entanglement \cite{buzek2} in detail.
\subsection{Optimal broadcasting of entanglement via local cloning}
\label{subsec:broad_ent_local}
In this subsection, we deal with the problem of broadcasting of quantum entanglement by using local cloning transformation.

The local output states $\tilde{\rho}_{13}$ on A's side and $\tilde{\rho}_{24}$ on B's side respectively and are given in canonical representation by, 
\begin{eqnarray}
\label{eq:gen_local_localoutputs1}
\tilde{\rho}_{13} = \left\{\frac{2}{3}\vec{x}, \frac{2}{3}\vec{x}, \frac{1}{3}\mathbb{I}_3 \right\},\hspace{0.1cm}\mbox{\&}\hspace{0.1cm} 
\tilde{\rho}_{24} = \left\{\frac{2}{3}\vec{y}, \frac{2}{3}\vec{y}, \frac{1}{3}\mathbb{I}_3 \right\},
\end{eqnarray}
where $\vec{x}$, $\vec{y}$ are the Bloch vectors of the initial state $\rho_{12}$. 

Next, we apply Peres-Horodecki criterion to investigate whether these local output states on either side of these two parties are separable or not. After evaluating determinats $W_2$, $W_3$ and $W_4$ (as given in Eq.~\eqref{eq:w3-w4}) we obtain a range involving input state parameters within which the local outputs, $\tilde{\rho}_{13}$ and $\tilde{\rho}_{24}$, are separable. These ranges for  $\tilde{\rho}_{13}$ and $\tilde{\rho}_{24}$ are
\begin{eqnarray}
0&\leq&\|\vec{x}\|\leq\frac{3}{4} \: \text{\& }\:\|\vec{x}\|\leq1+x_3+x_{3}^{2},\nonumber\\
0&\leq&\|\vec{y}\|\leq\frac{3}{4} \: \text{ \& } \: \|\vec{y}\|\leq1+y_3+y_{3}^{2}
\label{eq:range_local_separability}
\end{eqnarray}
respectively. Here $\|\vec{a}\|=\tr \left(a^{\dagger} a\right)$ with $\dagger$ denoting the Hermitian conjugate.

We have the nonlocal output states $\tilde{\rho}_{14}$ and $\tilde{\rho}_{23}$ as
\begin{eqnarray}
\tilde{\rho}_{14} = \tilde{\rho}_{23} =\left\{\frac{2}{3}\vec{x}, \frac{2}{3}\vec{y}, \frac{4}{9}T \right\},
\label{eq:gen_local_nonlocaloutputs1}
\end{eqnarray}
where $\vec{x}$, $\vec{y}$ are the Bloch vectors and $T$ is the correlation matrix of the initial state $\rho_{12}$.

Again with the help of Peres-Horodecki criterion we find out the condition under which the nonlocal output states will be inseparable. This condition for inseparability of the states $\tilde{\rho}_{14}$ and $\tilde{\rho}_{23}$ involving input state parameters is given as,
\begin{eqnarray}
\label{eq:inseparability_local_most_gen}
\left(W^l_3<0 \text{ or }  W^l_4 <0\right) \text{ and } W^l_2 \geqslant 0.
\end{eqnarray}
Here the explicit expressions of $W^l_2$, $W^l_3$ and $W^l_4$ are given by Eqs.~\eqref{eq:ws_local_inseparable_w2}, ~\eqref{eq:ws_local_inseparable_w3} and ~\eqref{eq:ws_local_inseparable_w4} in Appendix-1.

Now combining these two ranges determining the separability of the local states given by Eq.~\eqref{eq:range_local_separability} and inseparability of the nonlocal states given by Eq.~\eqref{eq:inseparability_local_most_gen}, we obtain the range for broadcasting of entanglement.

To exemplify our above study with a local cloner,  we next consider two different classes of mixed entangled states, namely: (a) werner-like states \cite{werner, wernerlike} and (b) Bell-diagonal states \cite{horodecki-entanglement, lang} and then separately analyse their broadcasting ranges.
\subsubsection{Example 2.1: Werner-like States}
First of all, we consider the example of werner-like states. These states can more formally be expressed as,
\begin{eqnarray}
\rho^w_{12} = \left\{\vec{x}^w, \vec{x}^w, T^w \right\}, 
\label{eq:wernerlike_state}
\end{eqnarray}
where $\vec{x}^w = \left\{ 0,\:  0,\:  p \left(\alpha^2-\beta^2\right)\right\}$ is the Bloch vector and the correlation matrix is $T^w=\di(2p\alpha\beta,-2p\alpha\beta,p)$ 
with the condition $\alpha^2+\beta^2=1$ and $0\leqslant p \leqslant1$. (Please note that whenever we use $M=\di(.,.,.)$, we mean $M$ is a diagonal matrix with diagonal elements given inside the first bracket.)

The local output states obtained after applying cloning operation on both the qubits $1$ and $2$ are given by,
\begin{eqnarray}
\tilde{\rho}_{13}=\tilde{\rho}_{24}=\left\{\frac{2}{3}\vec{x}^w,\: \frac{2}{3}\vec{x}^w,\: \frac{1}{3} \mathbb{I}_{3} \right\},
\end{eqnarray} 
where $\vec{x}^w$ is the Bloch vector of the state $\rho^w_{12}$.

From Peres-Horodecki theorem, if follows that by using Eq.~\eqref{eq:w3-w4} the local output states will be separable if either of the following two conditions are satisfied,
\begin{eqnarray}
0 &\leqslant& p \leqslant \frac{\sqrt{3}}{2}\: \text{\&}\: 0\leqslant \alpha^2 \leqslant 1, \:\mbox{Or,}\nonumber\\
\frac{\sqrt{3}}{2}< p &\leqslant& 1 \: \text{\&}\:  \frac{2p-\sqrt{3}}{4p}\leqslant \alpha^2 \leqslant \frac{\sqrt{3}+2p}{4p}.\label{eq:wernerlike_local_separable_zone}
\end{eqnarray}

Similarly after cloning, we have the nonlocal output states as,
\begin{eqnarray}
&\tilde{\rho}_{14}=\tilde{\rho}_{23}= \left\{\frac{2}{3}\vec{x}^w,\: \frac{2}{3}\vec{x}^w,\: \frac{4}{9}T^w \right\},
\end{eqnarray} 
where $\vec{x}^w$ is Bloch vector and $T^w$ is the correlation matrix of the state $\rho^w_{12}$.

\begin{figure}[h]
\begin{center}
\[
\begin{array}{cc}
\includegraphics[height=6cm,width=9cm]{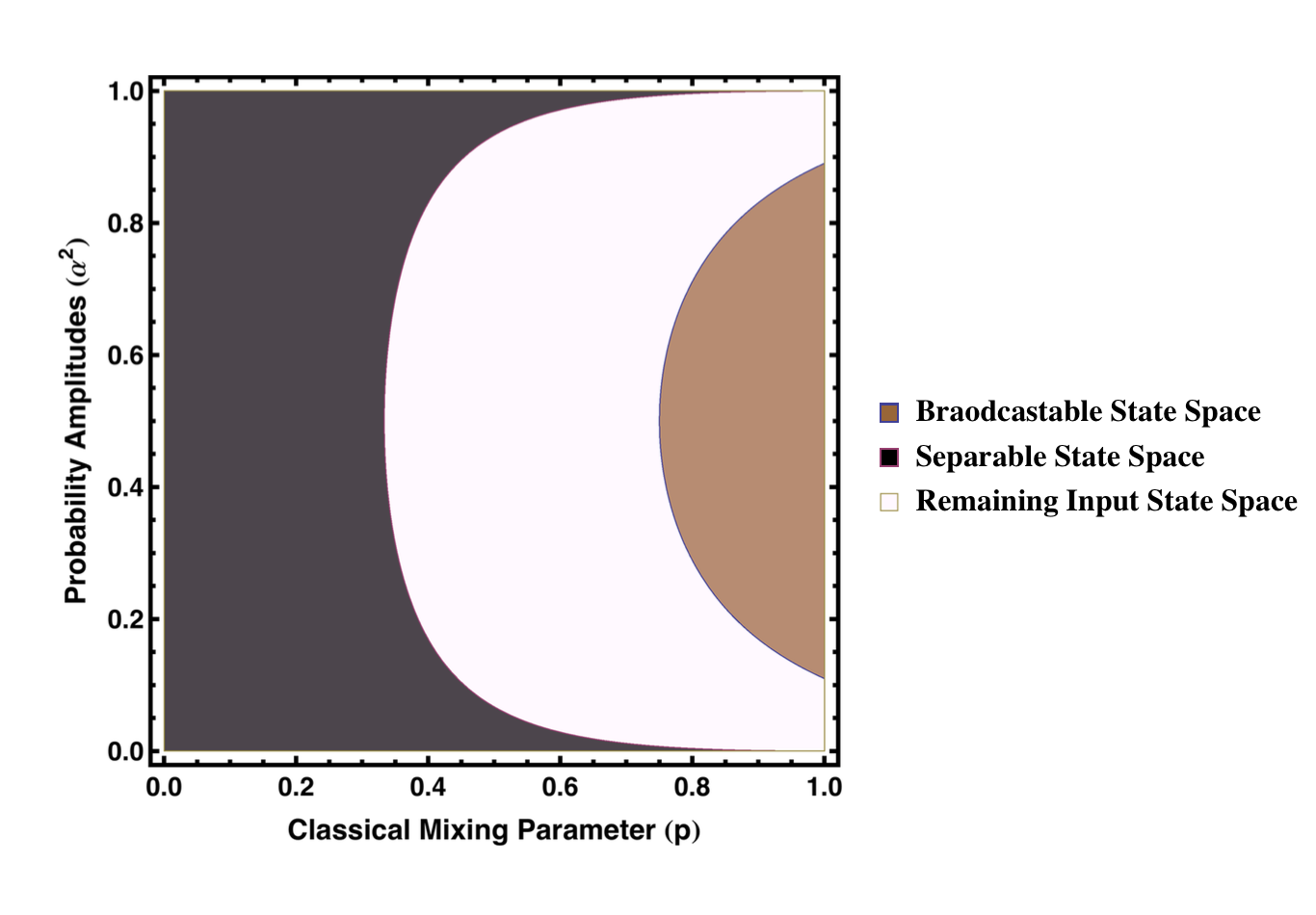}
\end{array}
\]
\end{center}
\caption{\noindent 
The figure illustates the states which can be used for broadcasting of entanglement via local cloning out of the total input state space of werner-like states $\rho^w_{12}$.\label{fig:werner_local}
}
\end{figure}

Using Peres-Horodecki theorem, the inseparability range of these nonlocal output states turn out to be,
\begin{equation}
\frac{3}{4}<p\leq1\, \text{ \& }\,N_-<\alpha^2 < N_+,
\label{eq:werner_broadrange_local}
\end{equation}
where $N_{\pm}=\frac{1}{16}\{8\pm(48-\frac{81}{p^2}+\frac{72}{p})^{\frac{1}{2}}\}$. On merging this inseparable zone along with the separable zone given by Eq.~\eqref{eq:wernerlike_local_separable_zone} we discover that the broadcasting range is exactly same as the inseparability range given by Eq.~\eqref{eq:werner_broadrange_local}. In FIG.~\ref{fig:werner_local}, we depict this broadcastable zone (given by Eq.~\eqref{eq:werner_broadrange_local}) among the allowed region of input state parameters $p$ and $\alpha$.
Next we provide two different tables for detailed analysis of the above broadcasting range. In TABLE~\ref{tbl:werner_like_local_1}, we give the broadcasting range of the werner-like states in terms of $p$ for different values of the input state parameter $\alpha^2$ and in terms of $\alpha^2$ for different values of the classical mixing parameter $p$.
\begin{table}[h]
\centering
\makebox[0pt][c]{\parbox{.5\textwidth}{%
    \begin{minipage}[b]{0.40\hsize}\centering
        \begin{tabular}{ | c | c |}
           \hline
           & Broadcasting\\$\alpha^{2}$ &  Range\\ \hline
           $0.2$ & $0.87<p\leqslant1$\\
           \hline 
           $0.4$ & $0.76<p\leqslant1$\\
           \hline
           $0.5$ & $0.75<p\leqslant1$\\
           \hline 
           $0.6$ & $0.76<p\leqslant1$\\
           \hline 
           $0.8$ & $0.87<p\leqslant1$\\
           \hline 
        \end{tabular}\\
(i)
    \end{minipage}
    \hfill
    \begin{minipage}[b]{0.50\hsize}\centering
        \begin{tabular}{ | c | c |}
           \hline
           & Broadcasting  \tabularnewline
           $p$ & Range\tabularnewline
           \hline 
           $0.76$ & $0.40<\alpha^2<0.60$ \tabularnewline
           \hline
           $0.85$ & $0.22<\alpha^2<0.78$\tabularnewline
           \hline 
           $0.9$ & $0.17<\alpha^2<0.83$ \tabularnewline
           \hline 
           $0.95$ & $0.14<\alpha^2<0.87$ \tabularnewline
           \hline 
           $1$ & $0.11< \alpha^2 <0.89$ \tabularnewline
           \hline 
        \end{tabular}\\
(ii)
    \end{minipage}
}}
\caption{Broadcasting ranges obtained using local cloners (i) in terms of $p$ for different values of $\alpha^2$ and (ii) in terms of $\alpha^2$ for different values of $p$.}
\label{tbl:werner_like_local_1}
\end{table}


\noindent \emph{Note 1}: We note that for $p=1$, Eq.~\eqref{eq:wernerlike_state} reduces to a non-maximally entangled state, for which the range for broadcasting of entanglement comes out to be \cite{buzek2}, $L_-<\alpha^{2}<L_+$, where $L_{\pm}=\frac{1}{16}(8\pm\sqrt{39})$.\\
\emph{Note 2}: Similarly we note that for $\alpha=\beta=\frac{1}{\sqrt{2}}$ (i.e. when $\left|\varphi\right\rangle _{12}$ is maximally entangled), Eq.~\eqref{eq:wernerlike_state} reduces to the Werner state \cite{werner}, for which the range for broadcasting of entanglement becomes, 
$\frac{3}{4}<p\leq1$.
\subsubsection{Example 2.2: Bell-diagonal States}
\label{subsubsec:ent_bell_local}
Here our initial resources are Bell-diagonal states to the local cloner which can be formally expressed as,
\begin{eqnarray}
\rho^b_{12} = \left\{\vec{0}, \vec{0}, T^b\right\},
\label{eq:belldiagonal_state}
\end{eqnarray}
where $\vec{0}$ is the Bloch vector which is a null vector and the correlation matrix is $T^b=
\di(c_1,c_2,c_3)$
with $-1\leqslant \text{c}_{i} \leqslant1$. 

The above input Bell-diagonal state can be rewritten as \cite{horodecki-entanglement, lang},
$\rho^b_{12}=\sum_{m,n} \lambda_{mn} \left|\gamma_{mn}\right\rangle\left\langle\gamma_{mn}\right|$
where the four Bell states  $\left|\gamma_{mn}\right\rangle\equiv\left(\left|0,n\right\rangle+(-1)^m \left|1,1\oplus n \right\rangle\right)/\sqrt{2}$ represents the eigenstates of $\rho^b_{12}$ with eigenvalues,
\begin{equation}
\lambda_{mn}=\frac{1}{4}\left[ 1+ (-1)^m c_1-(-1)^{(m+n)} c_2 + (-1)^n c_3\right].\nonumber 
\label{eq:eigenvalues_belldiag}
\end{equation}
Also, for $\rho^b_{12}$ to be a valid density operator, its eigenvalues have to be positive, i.e. $\lambda_{mn}\geqslant 0$.

Once again by applying local cloning and tracing out the qubits we get the local output states as:
\begin{eqnarray}
\tilde{\rho}_{13}=\tilde{\rho}_{24}=\left\{\vec{0}, \vec{0}, \frac{1}{3}\mathbb{I}_{3}\right\}.
\end{eqnarray}
It turns out that for these local output states both $W_3$ as well as $W_4$ given by Eq.~\eqref{eq:w3-w4} are non-negative and independent of the input state parameters ($c_i$'s). Hence, $\tilde{\rho}_{13}$ and $\tilde{\rho}_{24}$ will always remain separable.

On the other hand, the nonlocal outputs are given by,
\begin{eqnarray}
\tilde{\rho}_{14}=\tilde{\rho}_{23}=\left\{\vec{0}, \vec{0}, \frac{4}{9}T^b\right\},
\end{eqnarray}
where $T^b$ is the correlation matrix of the state $\rho^b_{12}$.

The inseparability range for these nonlocal output states of the input Bell-diagonal state $\rho^b_{12}$ in terms of $c_i$'s, is given by
\begin{eqnarray}
 -1\leq c_1 <-\frac{1}{4}\: \text{\&}\: \left(\gamma<-\frac{9}{4}  \: \text{or}\: \frac{9}{2}-c_-  <c_2\leq 1\right)\nonumber\\ \text{Or,}\:  \frac{1}{4}< c_1\leq 1\: \text{\&}\:  \left(c_-<c_2 \leq 1\: \text{or}\: -1\leq c_2<c_+\right),
\label{eq:bell_broadrange_local}
\end{eqnarray}
along with the condition that $\lambda_{mn} \geqslant 0$, where $c_{\pm}=\mp\frac{9}{4}\pm(c_1\pm c_3)$ and $\gamma=\tr(T^b)$. It is evident that the broadcasting range of the Bell-diagonal state is same as the inseparability range in Eq.~\eqref{eq:bell_broadrange_local} since the local output states in this case are always separable.

\begin{figure}[h]
\begin{center}
\[
\begin{array}{cc}
\includegraphics[height=8.5cm,width=8.5cm]{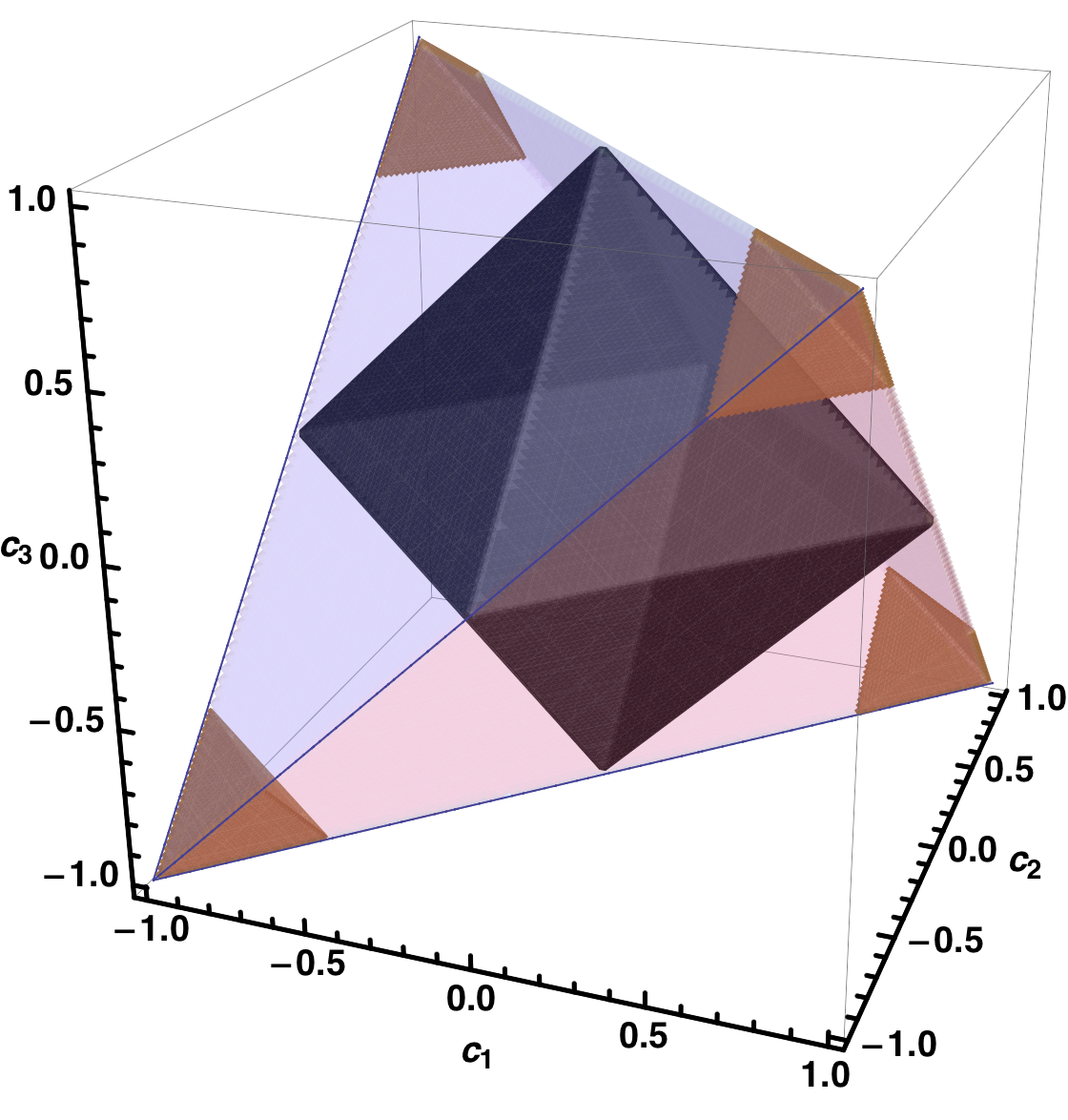}
\end{array}
\]
\end{center}
\caption{\noindent 
The figure illustates the broadcastable region obtained using local cloning operations within the geometry of Bell-diagonal states $\rho^b_{12}$. The translucent tetrahedron $\mathscr{T}$ hosts the Bell states $\left|\gamma_{mn}\right\rangle$ at the vertex tuples (-1,-1,-1), (1,1,-1), (1,-1,1) and (-1,1,1) from each of which a (brown) cone $\mathscr{C}$ emerges marking the broadcastable zones. The (black) octahedron $\mathscr{O}$ in the middle of the tetrahedron $\mathscr{T}$ depicts the separable region within the Bell-diagonal state space. \label{fig:bell_local}
}
\end{figure}

In FIG.~\ref{fig:bell_local}, we depict the above broadcastable zone (given by Eq.~\eqref{eq:bell_broadrange_local}) within the permissible region of the input state parameters, specified by the 3-tuple ($c_1$, $c_2$, $c_3$) from Eq.~\eqref{eq:belldiagonal_state}. Now for $-1\leqslant c_i \leqslant 1$, where $i$ = $\{1,2,3\}$, the condition that $\rho_{12}$ is necessarily a positive operator, i.e. $\lambda_{mn} \geqslant 0$, results in giving a tetrahedral geometrical representation of Bell-diagonal states $\mathscr{T}$ whose four vertices are the four Bell states or the eigenstates $\left|\gamma_{mn}\right\rangle$. The separable part within the geometry of Bell-diagonal states $\mathscr{T}$ comes out to be an octahedron $\mathscr{O}$ which is specified by the relation $|c_1|+|c_2|+|c_3|\leqslant1$ or $\lambda_{mn}\leqslant \frac{1}{2}$. Within the tetrahedron $\mathscr{T}$, the four entangled (inseparable) zones lie outside the octahedron $\mathscr{O}$, one from each vertex of $\mathscr{T}$ with the value 
of $\lambda_{mn}$ being greatest at the vertex points for each of them \cite{lang}. Interestingly, we discover that the broadcastable zone procured by using the above broadcasting condition in Eq.~\eqref{eq:bell_broadrange_local} turns out to be cones $\mathscr{C}$s, fitting as small caps on these entangled zones of the tetrahedron $\mathscr{T}$. It is also consistent with the fact that the maximally entangled states $\left|\gamma_{mn}\right\rangle$ lie at the vertices of $\mathscr{T}$, so the broadcastable regions start from those and vanish on the way towards the separable part $\mathscr{O}$. This is because the amount of entanglement keeps decreasing in the same direction. In other words, the states beyond the conic regions ($\mathscr{C}$s) lack the amount of initial entanglement required to be able to broadcast the same by local cloning operations.
It is interesting to observe that if $c_i=-1$ then $c_j=c_k$ and if $c_i=1$ then $c_j=-c_k$ where for each case $-1\leqslant c_j\:(c_k) < -\frac{5}{8}$ or $\frac{5}{8}< c_j \:(c_k) \leqslant 1$ with $i \neq j \neq k$ and $i,j,k=\{1,2,3\}$. This happens due to the symmetry of the Bell-diagonal states and that of the conic broadcasting zones as depicted in FIG.~\ref{fig:bell_local}. For the same reason, we also find that the four $\mathscr{C}$s or the conic zones grow symmetrically and uniformly from $c_i$'s = $-1$ ($1$) and ceases to exist for any value equal or beyond $-\frac{5}{8}$ ($\frac{5}{8}$). Hence in the TABLE~\ref{tbl:bell_diagonal_local}, we give the broadcasting range of Bell-diagonal states $\rho^b_{12}$ for different values of the first two input state parameters $c_1$, $c_2$ and variable over the third $c_3$, between the valid zone from $-1$ to $-5/8$ or $\frac{5}{8}$ to $1$. In this table, we restrict our results only to the negative range of inputs for $c_1$ and $c_2$ as the result of the 
broadcasting range in terms of $c_3$ remains unchanged when corresponding positive values of $c_1$ and $c_2$ are substituted in Eq.~\eqref{eq:bell_broadrange_local}.
\begin{center}
\begin{table}[!ht]
\begin{tabular}{|c|c|c|c|}
\hline 
$c_1$ & $c_2$ & Broadcasting Range\tabularnewline
\hline 
\hline 
$-\frac{7}{8}$ & $-\frac{7}{8}$ & $-1\leq c_3 \leq -\frac{3}{4}$\tabularnewline
\hline 
$-\frac{3}{4}$ & $-\frac{3}{4}$ & $-1\leq c_3 <-\frac{3}{4}$\tabularnewline
\hline 
$-\frac{7}{8}$ & $-\frac{3}{4}$ & $-\frac{7}{8}\leq c_3 <-\frac{5}{8}$\tabularnewline
\hline 
$-\frac{3}{4}$ & $-\frac{7}{8}$ & $-\frac{7}{8}\leq c_3 <-\frac{5}{8}$\tabularnewline
\hline 
\end{tabular}
\caption{\noindent 
Broadcasting ranges obtained with local cloners in terms of $c_3$ for different valid values of $c_1$ and $c_2$.}
\label{tbl:bell_diagonal_local}
\end{table}
\end{center}
\subsection{Optimal broadcasting of entanglement via nonlocal cloning}
\label{subsec:broad_ent_nonlocal}
In this subsection, we reconsider the problem of broadcasting of entanglement but this time by using nonlocal cloning transformation. 

The obtained nonlocal output states $\tilde{\rho}_{12}$ and $\tilde{\rho}_{34}$ are identical and they can be  represented as,
\begin{eqnarray}
&&\tilde{\rho}_{12} =\tilde{\rho}_{34} = \left\{\frac{3}{5}\vec{x}, \frac{3}{5}\vec{y}, \frac{3}{5}T \right\}
\label{eq:gen_nonlocal_nonlocaloutputs1}
\end{eqnarray}
where $\vec{x}$, $\vec{y}$ are the Bloch vectors and $T$ is the correlation matrix of the state $\rho_{12}$.

We apply the Peres-Horodecki criteria to find out the condition on input state parameters under which the above output states ($\tilde{\rho}_{12}$ and $\tilde{\rho}_{34}$) will be inseparable. This condition of inseparability turns out to be,
\begin{eqnarray}
W^{nl}_3<0 \text{ or }  W^{nl}_4 <0 \text{ \& } W^{nl}_2 \geqslant 0,
\label{eq:range_nonlocal_inseparability}
\end{eqnarray}
where the explicit expressions of $W^{nl}_2$, $W^{nl}_3$ and $W^{nl}_4$ are given by Eqs.~\eqref{eq:ws_nonlocal_inseparable_w2}, ~\eqref{eq:ws_nonlocal_inseparable_w3} and ~\eqref{eq:ws_nonlocal_inseparable_w4} in Appendix-2.

Next, the remaining states $\tilde{\rho}_{13}$ and $\tilde{\rho}_{24}$ are given by,
\begin{eqnarray}
\label{eq:gen_nonlocal_localoutputs1}
&\tilde{\rho}_{13} = \left\{\frac{3}{5}\vec{x}, \frac{3}{5}\vec{x}, \frac{1}{5}\mathbb{I}_{3} \right\},\hspace{0.1cm}\mbox{\&}\hspace{0.1cm}
\tilde{\rho}_{24} = \left\{\frac{3}{5}\vec{y}, \frac{3}{5}\vec{y}, \frac{1}{5}\mathbb{I}_{3} \right\}
\end{eqnarray}
where, $\vec{x}$ and $\vec{y}$ are the Bloch vectors of the state $\rho_{12}$. 

Similarly, here also we apply the Peres-Horodecki criterion to see whether these output states are separable or not. After evaluating determinants $W_2$, $W_3$ and $W_4$ (as given in Eq.~\eqref{eq:w3-w4}) we obtain a range involving input state parameters for which the output states, $\tilde{\rho}_{13}$ and $\tilde{\rho}_{24}$, are separable. This range is given by, 
\begin{eqnarray}
&& 0\leq\|\vec{x}\|\leq\frac{8}{9} \: \text{ \& }\: \|\vec{x}\| - x_{3}^{2} \leq \frac{4}{3}(1+x_3),  \nonumber\\
&&0\leq\|\vec{y}\|\leq\frac{8}{9} \: \text{ \& } \: \|\vec{y}\| - y_{3}^{2} \leq \frac{4}{3}(1+y_3)
\label{eq:range_nonlocal_separability}
\end{eqnarray} respectively.

Now, clubbing the two ranges given by Eq.~\eqref{eq:range_nonlocal_inseparability} and Eq.~\eqref{eq:range_nonlocal_separability}, we obtain the range for broadcasting of entanglement for $\rho_{12}$ via nonlocal copying. 

Next, in order to exemplify our study with nonlocal cloner we look into the broadcasting ranges of two different classes of input states: (a) Werner-like states \cite{werner, wernerlike} and (b) Bell-diagonal states \cite{horodecki-entanglement, lang}. 
\subsubsection{Example 3.1: Werner-Like State}
Quite similar to the previous section, here we reconsider the class of werner-like states given earlier by Eq.~\eqref{eq:wernerlike_state} and apply nonlocal cloning operation on it.

After cloning, the desired output states are given by,
\begin{eqnarray}
\tilde{\rho}_{12}=\tilde{\rho}_{34}= \left\{\frac{3}{5}\vec{x}^w,\: \frac{3}{5}\vec{x}^w,\: \frac{3}{5}T^w \right\},
\end{eqnarray}
where, $\vec{x}^w$ is the Bloch vector and $T^w$ is the correlation matrix of the state $\rho^w_{12}$.
The inseparability range for these states is given by,
\begin{equation}
\frac{5}{9}<p\leqslant1\, \text{ and }\,H_-<\alpha^2 < H_+,
\label{eq:wernerlike_broadrange_nonlocal}
\end{equation}
where $H_{\pm}=\frac{1}{2}\pm\{\frac{1}{144p}(27p^2+30p-25)\}^{\frac{1}{2}}$. The remaining output states are given by,
\begin{eqnarray}
\tilde{\rho}_{13}=\tilde{\rho}_{24}=\left\{\frac{3}{5}\vec{x}^w,\: \frac{3}{5}\vec{x}^w,\: \frac{1}{5}\mathbb{I}_3 \right\},
\end{eqnarray}
where $\vec{x}^w$ is the Bloch vector of the state $\rho^w_{12}$. These output states will be separable if either of the following two conditions are satisfied,
\begin{eqnarray}
0\leqslant p\leqslant d \:
\& \:(0&\leqslant&\alpha^2 \leqslant\xi_-,\:\text{or}\:\xi_+< \alpha^2 \leqslant 1),\nonumber\\
\text{Or,}\: 0 &\leqslant& p\leqslant 1 \: \text{\&}\:  \xi_-<\alpha^2 \leqslant \xi_+,\label{eq:werner_like_nonlocal_separability}
\end{eqnarray}
where $d=\sqrt{\frac{8}{9(1-2\alpha^2)^2}}$ $\xi_{\pm}=\frac{1}{6}(3\pm2\sqrt{2})$

After merging the separability and inseparability conditions given by Eq.~\eqref{eq:werner_like_nonlocal_separability} and Eq.~\eqref{eq:wernerlike_broadrange_nonlocal} respectively, the broadcasting range of the werner-like state turns out to be same as the inseparability range and is thus given by Eq.~\eqref{eq:wernerlike_broadrange_nonlocal}.

\begin{figure}[h]
\begin{center}
\[
\begin{array}{cc}
\includegraphics[height=6cm,width=9cm]{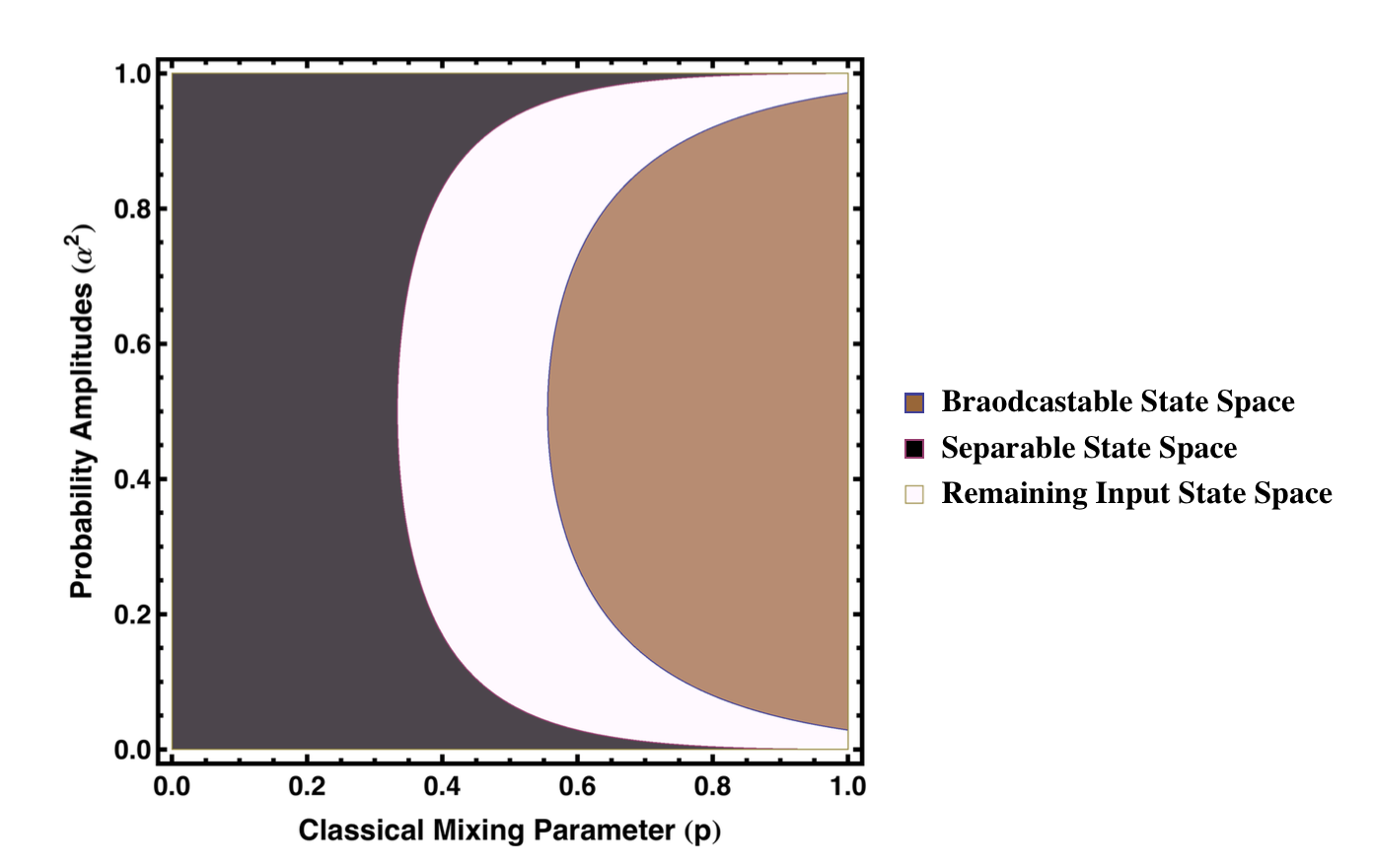}
\end{array}
\]
\end{center}
\caption{\noindent 
The figure illustates the states which can be used for broadcasting of entanglement via nonlocal cloning out of the total input state space of werner-like states $\rho^w_{12}$.\label{fig:werner_nonlocal}
}
\end{figure}

In FIG.~\ref{fig:werner_nonlocal}, we demarcate this broadcastable zone, given by Eq.~\eqref{eq:wernerlike_broadrange_nonlocal}, amidst the prescribed region of input state space.
Quite similar to the local cloning situation here also we provide two different tables for detailed analysis of the broadcasting range. In TABLE~\ref{tbl:werner_like_nonlocal_1}, we give the broadcasting range in terms of the classical mixing parameter $p$ for given values of input state parameter $\alpha^2$ and in terms of the input state parameter $\alpha^2$ for given values of classical mixing parameter $p$.
\begin{table}[h]
\centering
\makebox[0pt][c]{\parbox{.5\textwidth}{%
    \begin{minipage}[b]{0.40\hsize}\centering
        \begin{tabular}{ | c | c |}
           \hline
           & Broadcasting\\$\alpha^{2}$ &  Range\\ \hline
           $0.2$ & $0.64<p\leqslant1$\\
           \hline 
           $0.4$ & $0.56<p\leqslant1$\\
           \hline
           $0.5$ & $0.55<p\leqslant1$\\
           \hline 
           $0.6$ & $0.56<p\leqslant1$\\
           \hline 
           $0.8$ & $0.64<p\leqslant1$\\
           \hline 
        \end{tabular}\\
(i)
    \end{minipage}
    \hfill
    \begin{minipage}[b]{0.50\hsize}\centering
        \begin{tabular}{ | c | c |}
           \hline
           & Broadcasting  \tabularnewline
           $p$ & Range\tabularnewline
           \hline 
           $0.56$ & $0.42<\alpha^2<0.58$ \tabularnewline
           \hline 
           $0.65$ & $0.19<\alpha^2<0.81$\tabularnewline
           \hline 
           $0.85$ & $0.06<\alpha^2<0.94$ \tabularnewline
           \hline 
           $0.95$ & $0.04< \alpha^2< 0.96$ \tabularnewline
           \hline 
           $1$ & $0.03< \alpha^2 < 0.97$ \tabularnewline
           \hline 
        \end{tabular}\\
(ii)
    \end{minipage}
}}
\caption{Broadcasting ranges obtained using nonlocal cloners (i) in terms of $p$ for different values of $\alpha^2$ and (ii) in terms of $\alpha^2$ for different values of $p$.}
\label{tbl:werner_like_nonlocal_1}
\end{table}

\noindent \emph{Note 3:} We note that for $p=1$ case Eq.~\eqref{eq:wernerlike_state} reduces to a non-maximally entangled state, for which the range for broadcasting of entanglement comes out to be \cite{buzek3, kar},
 $\xi_-<\alpha^{2}<\xi_+.$\\
\emph{Note 4:} Again for $\alpha=\beta=\frac{1}{\sqrt{2}}$ (i.e. when $\left|\varphi\right\rangle _{12}$ is maximally entangled) Eq.~\eqref{eq:wernerlike_state} reduces to the Werner state \cite{werner}, for which the range for broadcasting of entanglement becomes,
 $\frac{5}{9}<p\leq1$.
\subsubsection{Example 3.2: Bell-diagonal states}
\label{subsubsec:ent_bell_local}
In this example, we once again consider the Bell-diagonal states (given earlier by Eq.~\eqref{eq:belldiagonal_state}) as our initial entangled state.

Once the nonlocal cloner is applied to it we have the desired output states as,
\begin{eqnarray}
\tilde{\rho}_{12}=\tilde{\rho}_{34}= \left\{\vec{0}, \vec{0}, \frac{3}{5}T^b\right\},
\end{eqnarray}
where $T^b$ is the the correlation matrix of the state $\rho^b_{12}$.

The inseparability range of the desired output states is given by,
\begin{eqnarray}
& (6c_1-3\gamma +5) (3\gamma-6c_3-5) (3\gamma-6 c_2-5)(3\gamma+ \nonumber\\
& 5)<0\:\text{or}\: (3 c_3+5) \left((5-3 c_3)^2-9 (c_1-c_2)^2\right)<0 \:\:
\label{eq:belldiag_broadrange_nonlocal}
\end{eqnarray}
where $\gamma=\tr(T^b)$ along with the condition that $\lambda_{mn} \geqslant 0$ from the positivity of input density operator $\rho_{12}$.

The remaining output states are given by,
\begin{eqnarray}
\tilde{\rho}_{13}\:=\:\tilde{\rho}_{24}= \left\{\vec{0}, \vec{0}, \frac{1}{5}\mathbb{I}_{3}\right\}.
\end{eqnarray}

These output states are independent of the input state parameter  ($c_i$'s) and will be always separable since for them the $W_3$ and $W_4$ from Eq.~\eqref{eq:w3-w4} comes out to be a positive number. Hence, the broadcasting range of the Bell-diagonal state is same as the inseparability range as given in Eq.~\eqref{eq:belldiag_broadrange_nonlocal}.

\begin{figure}[h]
\begin{center}
\[
\begin{array}{cc}
\includegraphics[height=8.5cm,width=8.5cm]{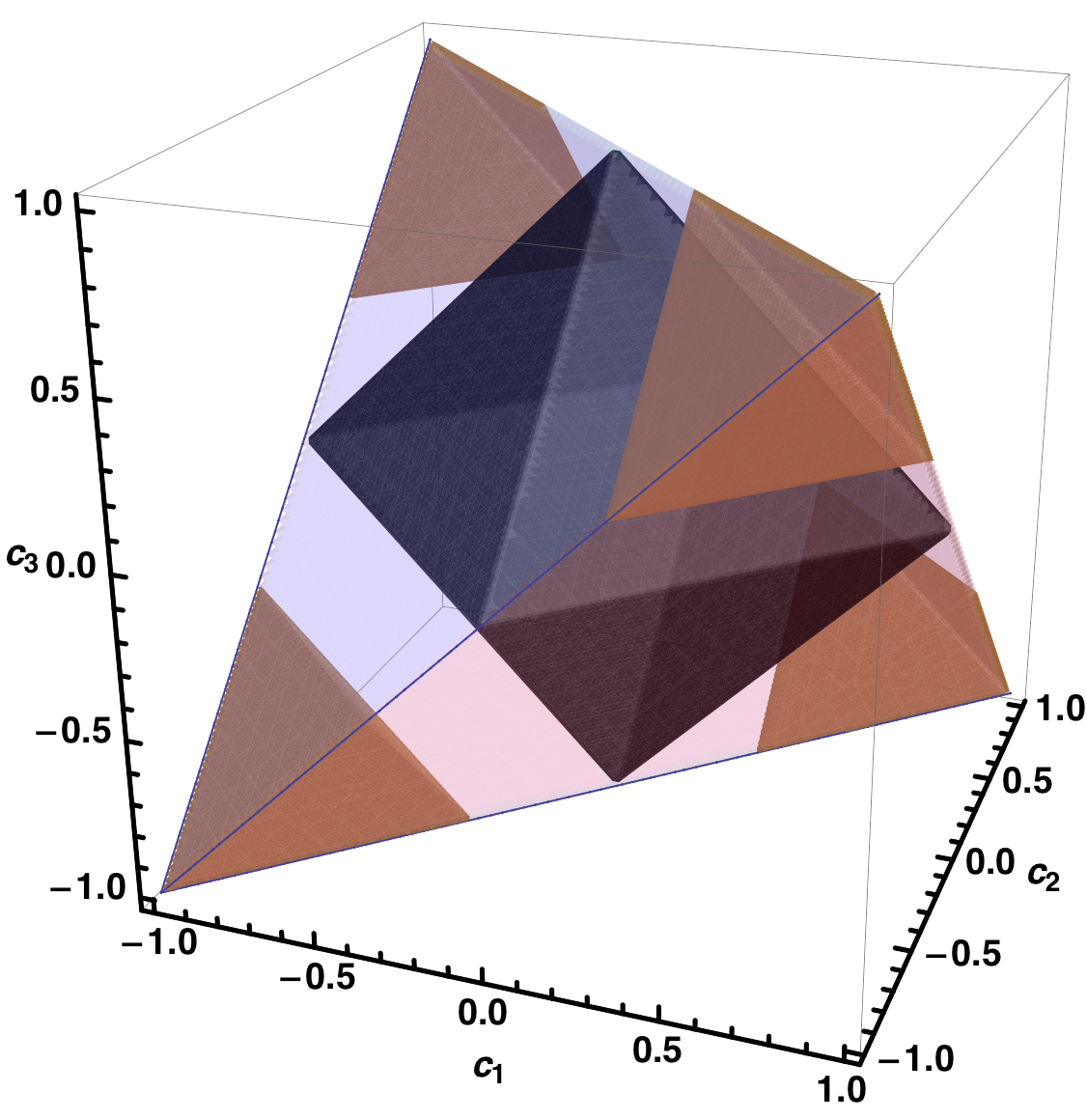}
\end{array}
\]
\end{center}
\caption{\noindent 
The figure illustates the broadcastable region obtained using nonlocal cloning operations within the geometry of Bell-diagonal states $\rho^b_{12}$. The translucent tetrahedron $\mathscr{T}$ hosts the Bell states $\left|\gamma_{mn}\right\rangle$ at the vertex tuples (-1,-1,-1), (1,1,-1), (1,-1,1) and (-1,1,1) from each of which a (brown) cone $\mathscr{C'}$ emerges marking the broadcastable zones. The (black) octahedron $\mathscr{O}$ in the middle of the tetrahedron $\mathscr{T}$ depicts the separable region within the Bell-diagonal state space. Interestingly enough, by the use of nonlocal cloner we find that the height broadcastable conic regions have increased considerably compared to that obtained in FIG.~\ref{fig:bell_local} with local cloners. \label{fig:bell_nonlocal}
}
\end{figure}

Quite analogous to our geometric analysis in local copying case of the broadcasting region of Bell-diagonal state, in FIG.~\ref{fig:bell_nonlocal}, we depict the above broadcastable zone (given by Eq.~\eqref{eq:belldiag_broadrange_nonlocal}) among the allowed region of the input state parameters, specified by the 3-tuple ($c_1$, $c_2$, $c_3$) from Eq.~\eqref{eq:belldiagonal_state}. 
Similarly as in the case with local cloners, here also we notice that if $c_i=-1$ then $c_j=c_k$ and if $c_i=1$ then $c_j=-c_k$ where for each case $-1\leqslant c_j\:(c_k) < -\frac{1}{3}$ or $\frac{1}{3}< c_j \:(c_k) \leqslant 1$ with $i \neq j \neq k$ and $i,j,k=\{1,2,3\}$. This happens due to the symmetry of the Bell-diagonal states and that of the conic broadcasting zones as depicted in FIG.~\ref{fig:bell_nonlocal}. For the same reason, we also find that the four $\mathscr{C}$s or the conic zones grow symmetrically and uniformly from $c_i$'s = $-1$ ($1$) and ceases to exist for any value equal or beyond $-\frac{1}{3}$ ($\frac{1}{3}$). Hence in TABLE~\ref{tbl:bell_diagonal_nonlocal}, we give the broadcasting range of Bell-diagonal states $\rho^b_{12}$ for different values of the first two input state parameters $c_1$, $c_2$ and variable over the third $c_3$, between the valid zone from $-1$ to $-\frac{1}{3}$ or $\frac{1}{3}$ to $1$. In this table, we restrict our results only to the negative range of 
inputs for $c_1$ and $c_2$ as the result of the broadcasting range in terms of $c_3$ remains unchanged when corresponding positive values of $c_1$ and $c_2$ are substituted in Eq.~\eqref{eq:bell_broadrange_local}.
\begin{center}
\begin{table}[!ht]
\begin{tabular}{|c|c|c|c|}
\hline 
$c_1$ & $c_2$ & Broadcasting Range\tabularnewline
\hline 
\hline 
$-\frac{7}{9}$ & $-\frac{7}{9}$ & $-1\leq c_3 \leq -\frac{5}{9}$\tabularnewline
\hline 
$-\frac{5}{9}$ & $-\frac{5}{9}$ & $-1\leq c_3 <-\frac{5}{9}$\tabularnewline
\hline 
$-\frac{7}{9}$ & $-\frac{5}{9}$ & $-\frac{7}{9}\leq c_3 <-\frac{1}{3}$\tabularnewline
\hline 
$-\frac{5}{9}$ & $-\frac{7}{9}$ & $-\frac{7}{9}\leq c_3 <-\frac{1}{3}$\tabularnewline
\hline 
\end{tabular}
\caption{\noindent 
Broadcasting ranges obtained with nonlocal cloners for different valid values of $c_1$ and $c_2$.}
\label{tbl:bell_diagonal_nonlocal}
\end{table}
\end{center}
Interestingly, here we find for the above two cases that the use of a nonlocal cloner despite being difficult to implement gives us a much wider broadcasting range for entanglement. 
In non-local cloning of entanglement, 
the bipartite system as a whole gets entangled with a single cloning machine, whereas in local cloning each individual subsystem separately gets entangled with a cloning machine. A larger amount of entanglement transfer to the machine takes place in the local cloning case. So indeed it is not surprising that nonlocal cloning will produce a wider range for broadcasting of entanglement than the local cloning \cite{kar}.
\section{Broadcasting of Quantum Correlations Beyond Entanglement}
\label{sec:broaddiscord}
In this section, we consider broadcasting of quantum correlations which go beyond the notion of entanglement. 
Here, we analyse the possibility of creating more number of lesser correlated quantum states from an intial quantum state having correlations using cloning operations. 
\subsection{Quantum correlations beyond entanglement}
\label{subsec:discord}
Though QCs is synonymous to entanglement for pure two qubit quantum states, however precise nature of the QCs is not well understood for two-qubit mixed states and multipartite states \cite{werner, pankaj}. It has been suggested that QCs go beyond the simple idea of entanglement \cite{knill} i.e., QCsbE. The basic idea of quantum discord and other measures are to quantify all types of QCs including entanglement \cite{discord, modi-review, dissension}. 
Physically, quantum discord captures the amount of mutual information in multipartite systems which are locally inaccessible \cite{zhang}. There is another approach to quantify QCsbE. This is done by distance based measures. Distance-based discord is defined as the minimal distance between a quantum state and all other states with zero discord \cite{dakic, discord-luo, girolami}. 
It is similar to the geometric measure of quantum entanglement \cite{entanglement-geometric}. As a result, this kind of measure is also called the geometric measure of quantum discord (or simply \emph{geometric discord}). Here, we use this particular measure of discord to quantify the amount of QCsbE present in between a pair of qubits although our results hold for any measures of discord (QCsbE). 

\noindent\emph{\textbf{Geometric Discord}}\cite{girolami}: \noindent Geometric discord (GD) or square norm-based discord \cite{dakic, discord-luo} of any general two qubit state $\rho_{12}$ (of the form given by Eq.~\eqref{eq:mix}) is defined as, 
\begin{equation}
D_{G}(\rho_{12})=\underset{\chi}{\min}\left\Vert\rho_{12}\right.-\chi\left\Vert\right.^{2},
\end{equation}
where the minimum is over all possible classical states $\chi$ which is of the form $p\left|\psi_{1}\right\rangle \left\langle \psi_{1}\right|\otimes\rho_{1}+(1-p)\left|\psi_{2}\right\rangle \left\langle \psi_{2}\right|\otimes\rho_{2}$. Here, $\left|\psi_{1}\right\rangle$ and $\left|\psi_{2}\right\rangle$ are two orthonormal basis of subsystems $A$. The states $\rho_{1}$ and $\rho_{2}$ are two density matrices of subsystem $B$. In the above equation, $\left\Vert\right.\rho_{12}-\chi\left\Vert\right.^{2}=\tr(\rho_{12}-\chi)^{2}$ is referred to as the square norm of the Hilbert-Schmidt space. For an arbitrary two-qubit system (given by Eq.~\eqref{eq:mix}), an analytical expression of GD has been obtained \cite{dakic}, which is
\begin{equation} \label{eq:geodiscord}
D_{G}(\rho_{12}) = \frac{1}{4}(\left\Vert\right.\vec{x}\left\Vert\right.^{2}-\left\Vert\right.T\left\Vert\right.^{2}-\lambda_{\max}), 
\end{equation} 
where $\lambda_{\max}$ is the maximal eigenvalue of matrix $\Omega$ (= $\vec{x}\vec{x}^{t}-TT^{t}$). 
Here the superscript \emph{t} stands for transpose of a vector or matrix.

It is well known that geometric discord (GD) defined above can increase under local unitary e.g., 
under a simple channel $\Lambda$: $\rho\rightarrow\rho\otimes\sigma$, i.e., a channel which introduces an ancilla only \cite{pianigd}. 
In order to overcome this, it was  suggested that we can use different distance measures (norms) which will overcome this 
shortcoming \cite{othernorm}. Although information theoretic discord \cite{discord, modi-review} and GD using trace distance norm are invariant under local unitary, in general QCsbE are not monotone under any local operations.   
According to Streltsov \textit{et al.} \cite{Streltsov}: \textit{A local quantum channel 
acting on a single qubit can create QCsbE in a multiqubit system if and only if it is not unital}.  

Hence, we  discuss the broadcasting of QCsbE under two types of channel a) unital channel ($\Lambda_u$): $\mathbb{I}\rightarrow\mathbb{I}$ and b) non-unital channel $\Lambda_{nu}$: $\mathbb{I}\nrightarrow\mathbb{I}$. We will call this type of operations on the bonafied states as `processing': `pre-pocessing' (applying the channel on the input state before broadcasting) or `post-processing' (applying the channel on the output states after broadcasting).   
\subsection{Definition of broadcasting of QCsbE via. local and nonlocal cloning operations}
\label{subsec:broadcast_corr}
Here, we define what we mean by the broadcasting of QCs by using state independent (optimal) and state dependent B-H cloning machines. These cloning machines are applied both locally and nonlocally.

The scenario of broadcasting of QCsbE is similar to that of broadcasting of entanglement (see Fig. (\ref{fig:local_ent} \& \ref{fig:nonlocal})). Let $Q$ be the total amount 
 of QCsbE produced as a result of both local or  non local cloning and the sum of 
 the QCsbE within parties ($Q_{l}$) and the QCsbE across 
 the parties ($Q_{nl}$) then $Q=Q_{l}+Q_{nl}$. To maximize $Q_{nl}$, we must have $Q_{l}=0$.  


\noindent\textbf{Definition 3.3.1:} 
\label{def:broad_local}
A quantum correlated state $\rho_{12}$ is said to be broadcast after the application of local cloning 
operation ($U_1 \otimes U_2$), if for some values of the input state parameters, the amount of QCsbE for the 
states \{$\tilde{\rho}_{14}$, $\tilde{\rho}_{23}$\} are non-vanishing.

\noindent\textbf{Definition 3.3.2:} 
\label{def:broad_local}
A quantum correlated state $\rho_{12}$ is said to be broadcast after the application of nonlocal cloning 
operation ($U_{12}$), if for some values of the input state parameters, the QCsbE for the 
states \{$\tilde{\rho}_{12}$, $\tilde{\rho}_{34}$\} are non-vanishing.

\noindent\textbf{Definition 3.3.3:} 
\label{def:broad_local}
A quantum correlated state $\rho_{12}$ is said to be optimally broadcast after the application of local cloning 
operation ($U_1 \otimes U_2$), if for some values of the input state parameters, the QCsbE for the 
states \{$\tilde{\rho}_{14}$, $\tilde{\rho}_{23}$\} are non-vanishing 
and for the states \{$\tilde{\rho}_{13}$, $\tilde{\rho}_{24}$\}, the amount of QCsbE are zero.

\noindent\textbf{Definition 3.3.4:} 
\label{def:broad_local}
A quantum correlated state $\rho_{12}$ is said to be optimally broadcast after the application of nonlocal cloning 
operation ($U_{12}$), if for some values of the input state parameters, the QCsbE for the 
states \{$\tilde{\rho}_{12}$, $\tilde{\rho}_{34}$\} are non-vanishing 
whereas for the states \{$\tilde{\rho}_{13}$, $\tilde{\rho}_{24}$\}, the QCsbE are zero.

\subsection{Optimal Broadcasting of QCsbE via. local and nonlocal cloning operations under unital channel ($\Lambda_u$)}
\label{subsec:broadcast_corr}
In this subsection, we investigate the problem of broadcasting of QCsbE by using state independent (optimal) and state dependent B-H cloning machines under the unital channel ($\Lambda_u$). These cloning machines are applied both locally and nonlocally. As QCsbE are non-incrasing under $\Lambda_u$, it is evident that we need not to mention it everytime.
\subsubsection{Broadcasting of correlations using Buzek-Hillery (B-H) local cloners}
\label{subsub:broad_corr_B-Hopt_local}
Here we use B-H state independent optimal ($U^l_{bhsi}$) and state dependent ($U^l_{bhsd}$) cloning operation locally (given by Eq.~\eqref{eq:B-H_gen_transform}) and we find that it is possible to broadcast QCsbE by such methods but contrary to the broadcasting of entanglement, we will not have optimal one.
\noindent \begin{theorem}
\label{theo:local_broad}
\textit{Given a two qubit general mixed state $\rho_{12}$ and B-H local cloning transformations (state independent optimal $U^l_{bhsi}$ or state dependent $U^l_{bhsd}$), it is impossible to broadcast the QCsbE optimally within $\rho_{12}$ into two lesser quantum correlated states: \{$\tilde{\rho}_{14}$, $\tilde{\rho}_{23}$\}.  
}
\end{theorem}

\noindent \emph{Proof:} When B-H state dependent cloning transformation $U^{l}_{bhsd}$ (given by Eq.~\eqref{eq:B-H_gen_transform}) is applied locally to clone the qubits `$1\rightarrow 3$' and `$2\rightarrow4$' of an input most general mixed quantum state $\rho_{12}$, then we have the local output states as, 
$\tilde{\rho}_{13}$ = $\{ \mu \vec{x}, \:\mu \vec{x},\:  T^{sd}_l \}$ and $\tilde{\rho}_{24}$ = $\{ \mu \vec{y},\: \mu \vec{y},\:  T^{sd}_{l} \}$; 
where $T^{sd}_l=\di(
2\lambda,2\lambda,1-4\lambda)
$
and the nonlocal output states, $\tilde{\rho}_{14}$ = $\tilde{\rho}_{23}$ = $\{ \mu \vec{x}, \:\mu \vec{y},\:  \mu\:  T \}$. 
Here $\mu=1-2\lambda$; $\vec{x}$ and $\vec{y}$ represent the Bloch vectors and $T$ denotes the correlation matrix of the input 
state $\rho_{12}$. The GD $D_G$, calculated using Eq.~\eqref{eq:geodiscord}, of the local output states are given by  
$D_G(\tilde{\rho}_{13})=\frac{1}{2} \left(1+\mu^2 \|\vec{x}\| - 8\lambda + 20\lambda^2 \right)$ and 
$D_G(\tilde{\rho}_{24})=\frac{1}{2} \left(1+ \mu^2 \|\vec{y}\| - 8\lambda + 20\lambda^2 \right)$ 
which always remains non-vanishing for $0\leqslant\lambda\leqslant\frac{1}{2}$. 
This is because the minima of $D_G(\tilde{\rho}_{13})$ and $D_G(\tilde{\rho}_{24})$ come out to be $D^{min}_G=\frac{w}{2}-\frac{2}{5}$ at $\lambda=\frac{1}{5}$; where $w=1+\mu^2 \|\vec{x}\| $ or $w=1+\mu^2 \|\vec{y}\| $, giving $w\geqslant1$ and ensuring always that $D^{min}_G>0$. 

Hence we will never have optimal broadcasting of QCsbE although it is possible that we can have task oriented one.
%
%
%
\subsubsection{Broadcasting of correlations using Buzek-Hillery (B-H) nonlocal cloners}
\label{subsub:broad_corr_B-Hopt}
In this approach, we use symmetric B-H state independent optimal ($U^{nl}_{bhsi}$) as well as state dependent ($U^{nl}_{bhsd}$) nonlocal cloning operations (given by Eq.~\eqref{eq:B-H_gen_transform}) and we find that, here also it is possible to broadcast QCsbE by such approaches but not the optimal one.
\noindent \begin{theorem}
\label{theo:nonlocal_broad}
\textit{Given a two qubit general mixed state $\rho_{12}$ and B-H nonlocal cloning transformations (state independent optimal $U^{nl}_{bhsi}$ or state dependent $U^{nl}_{bhsd}$), it is impossible to broadcast the QCsbE optimally within $\rho_{12}$ into two lesser quantum correlated states: \{$\tilde{\rho}_{12}$, $\tilde{\rho}_{34}$\}.
}
\end{theorem}
\noindent \emph{Proof:} 
When B-H state dependent nonlocal cloning transformation $U^{nl}_{bhsd}$ (given by Eq.~\eqref{eq:B-H_gen_transform}) is applied to clone the qubits $1\:\&\: 2$ of an input most general mixed two qubit state $\rho_{12}$ (given in Eq.~\eqref{eq:mix}), then we have the output states, 
$\tilde{\rho}_{13}$ = $\{ \mu \vec{x}, \:\mu \vec{x},\:  T^{sd}_{nl} \}$ and 
$\tilde{\rho}_{24}$ = $\{ \mu \vec{y},\: \mu \vec{y},\:  T^{sd}_{nl} \}$; 
where $T^{sd}_{nl}=\di(
2\lambda,2\lambda,1-8\lambda)
$ and the desired output states, $\tilde{\rho}_{12}$ = $\tilde{\rho}_{34}$ = $\{ \mu \vec{x}, \:\mu \vec{y},\:  \mu\:  T \}$; where $\mu=1-4\lambda$. Here $\vec{x}$ as well as $\vec{y}$ represent the Bloch vectors and $T$ denotes the correlation matrix of the input state. The GD $D_G$, calculated using Eq.~\eqref{eq:geodiscord}, of the local output states are given by: 
$D_G(\tilde{\rho}_{13})=\frac{1}{2} \left(1+ \mu^2 \|\vec{x}\| - 16\lambda + 68\lambda^2 \right)$ and 
$D_G(\tilde{\rho}_{24})=\frac{1}{2} \left(1+ \mu^2 \|\vec{y}\| - 16\lambda + 68\lambda^2 \right)$ 
which always remains non-vanishing for $0\leqslant\lambda\leqslant\frac{1}{4}$. 
This is because the minima of $D_G(\tilde{\rho}_{13})$ and $D_G(\tilde{\rho}_{24})$ come out to be $D^{min}_G=\frac{1+5w}{34+8w}$ at $\lambda=\frac{2+w}{17+4w}$; where $w=\|\vec{x}\| $ or $w=\|\vec{y}\| $, giving $0\leqslant w \leqslant1$ and ensuring always that $D^{min}_G>0$.
%
%
%
Hence we will never have optimal broadcasting of QCsbE although it is possible that we can have task oriented one.

Now moving beyond the realms of the above theorems, we claim that \textit{if in the case of B-H state independent optimal cloners, when applied locally or nonlocally, we are unable to broadcast the QCsbE optimally then no other state independent deterministic cloner can do so}. It is mainly because of the recent result by Sazim \textit{et al} that for a given input state, the outputs of an optimal cloner are least correlated since as the fidelity of cloning increases the correlations transfer to the machine state also grows \cite{sazim-cloning}. Again in 2003, Ghiu \textit{et al} showed that entanglement is optimally broadcast and maximal fidelities of the two final entangled states are obtained only when symmetric cloning machines are applied \cite{ghiu}. So by combining the above two results by Sazim \textit{et al} and Ghiu \textit{et al}, we can logically infer that \textit{even asymmetric Pauli cloning machines will be unable to broadcast QCsbE optimally since for those also the local outputs will always 
possess non-vanishing GD} \cite{sazim-cloning, ghiu}. This enables us to comprehensively conclude that optimal broadcasting of QCsbE for any two qubit state via cloning operations is impossible.
\subsection{Optimal Broadcasting of QCsbE via. local and nonlocal cloning operations under Nonunital channel ($\Lambda_{nu}$)}
In this subsection, we will discuss the possibilities and impossibilities of broadcasting QCsbE under non-unital 
channel ($\Lambda_{nu}$). Here many situations can occur depending on the free will of the parties: a) pre-possesing 
the state with unital channel \& post-processing with non-unital channel, b) pre-processing with non-unital channel \& 
post-processing with unital channel, and c) pre- \& post-procesing with nonunital channel. All these situations are 
equivalent in the sense that QCsbE can increase under $\Lambda_{nu}$. 

It is also evident that we can have task oriented broadcasting of QCsbE and can increase the QCsbE of the broadcasted states 
if needed. 
\textit{And conceptually the notion of optimal broadcasting of QCsbE is not clear as we can have quantum correlated broadcast states 
although we start with totally classical correlated states}. 
\section{Conclusion}
\label{sec:conclusion}
\noindent In literature, generalized approaches exist for purification or compression of entanglement procedures but no such generalization exists for broadcasting (decompression) of entanglement via cloning operations \cite{buzek2, horodecki-purification}. Such a study can aid in discovering operational meaning of quantifying the amount of entanglement \cite{entanglement-geometric}. In a nutshell, in this work we present a holistic picture of broadcasting of quantum entanglement via cloning from any input two qubit state. We explicitly provide a set of ranges in terms of input state parameters for a most general representation of two qubit states for which broadcasting of entanglement will be possible. We exemplify our generalized results by examining them for two class of states: (a) Werner-like and (b) Bell-diagonal. We perform this study with both type of cloning techniques, local and nonlocal, to examine how the range of broadcasting increases under nonlocal cloning operations. 
Thereafter, we focus on the question whether broadcasting of QCsbE via cloning operations is possible or not. Contrary to the broadcasting of entanglement, we find that it is impossible to broadcast such QCsbE optimally via  cloning operations, whether local or nonlocal, from a given quantum mechanically correlated pair to two lesser correlated pairs. But we can have task oriented broadcasting for QCsbE. 
We also explicitly reason out why the local outputs from cloner (state dependent or state independent) will never possess vanishing QCsbE which is imperative to broadcast QCsbE. However, we can intuitively conjecture that if one tries to broadcast QCsbE to more than two pairs, say $N$ pairs, from an initial two qubit state then for some $N>2$ pairs there is possibility of success in broadcasting such correlations optimally. This is because the nonlocal outputs become unentangled when $1\rightarrow 3$ and $1\rightarrow7$ pairs are generated by the optimal local and nonlocal cloners respectively, which hints that the QCsbE in the output states decreases as more pairs are produced by the cloner \cite{kar}.

Our findings brings out a fundamental difference between the correlation defined from the perspective of entanglement and the correlation measure which claims to go beyond entanglement. 

\noindent\textbf{Acknowledgment:} S. Chatterjee gratefully acknowledges Prof. S. Chaturvedi and Prof. G. Kar for many insightful discussions which helped immensely in carrying out various calculations. We would also like to thank the anonymous referee for useful comments which indeed immensely improve our manuscript. 

\begin{widetext}
\section*{Appendix-1: Inseparability range of nonlocal outputs obtained using local cloners}
\label{app:1}
\noindent In this part, we evaluate the determinants $W_2$, $W_3$ and $W_4$ (as given in Eq.~\eqref{eq:w3-w4}) of the Peres-Horodecki criterion for the states $\tilde{\rho}_{14}$ and $\tilde{\rho}_{23}$ given by Eq.~\eqref{eq:gen_local_nonlocaloutputs1}, and denote them as $W^l_2$, $W^l_3$ and $W^l_4$ respectively. The mathematical expressions of these determinants are given as follow,
\begin{eqnarray}
&& W^l_2=-\frac{1}{6^4}\left[4 \displaystyle\sum_{i=1}^3(-1)^{\delta_{3i}}\left(t_{3i}+3y_i\right)^2+9\left(2x_3+3\right)^2\right],
\label{eq:ws_local_inseparable_w2}
\end{eqnarray}
\begin{eqnarray}
&& W^l_3 = L_f + \frac{1}{3^6} \left[2\displaystyle \sum_{i,j}^{2}t_{ij}t_{i3}t_{3j} + t_{33} \left(\displaystyle\sum_{i=1}^{3}t^2_{i3}+\displaystyle\sum_{i=1}^{2}t^2_{3i} - \displaystyle \sum_{i,j}^{2}t^2_{ij}\right)-\frac{9}{4}\left\{t_{33}\displaystyle \sum_{i,j}^{3}t^2_{ij}+3\displaystyle \sum_{i=1}^3 \left(t_{i3}x_i+t_{3i}y_i\right)\right\} \right. \nonumber\\
&& \left.
-\frac{3}{2}\left\{ \displaystyle \sum_{j=1}^{3} g_3 \displaystyle\sum_i^2 \{t_{ij}^2 - x_3 t_{3j}^2-y_3t_{j3}^2\}-3\left(\displaystyle\sum_{i,j}^{3}t_{ij}x_iy_j  -\displaystyle \sum_{i=1}^{2}\left\{2t_{i3}x_iy_3-\displaystyle\sum_{j=1}^2\left(t_{ij}x_jy_i-t_{ii}x_jy_j\right)\right\}\right) \right\}\right.\nonumber\\
&& \left.
+3\displaystyle\sum_{i\neq j}^{2}\left\{(t_{ii}-t_{jj})(x_i t_{3i} +y_i t_{i3}) +(t_{ij}-t_{ji})(x_i t_{3j} +y_i t_{j3})+(x_i t_{i3} +y_i t_{3i})t_{33}\right\}\right],
\label{eq:ws_local_inseparable_w3}
\end{eqnarray}
\begin{eqnarray}
&& W^l_4 = \frac{1}{6^8}\left[K_2+6^4\displaystyle\sum_{i,j}^{3}\{4t_{ij}x_iy_j- t^2_{ij}\}+\frac{2}{9}\left\{\frac{1}{2}\displaystyle \sum_{i,j}^{3} t_{ij}^4-\displaystyle\sum_{i,j}^2\displaystyle\sum_{p=j+1,q=i+1}^3(t_{ij}^2t_{qp}^2-4t_{ij}t_{ip}t_{qj}t_{qp})\right.\right.\nonumber\\
&& \left.\left.+\displaystyle \sum_{j=1}^3\displaystyle\sum_{p=j+1,q}^{3,j}S_qt_{1j}^2t_{pq}^2+4\left\{\displaystyle \sum_{i,j}^{3}x_iy_j\displaystyle \sum_{p\neq i,j}^3 \left(t_{ji}t_{pp}-t_{jp}t_{pi}\right)-\displaystyle \sum_{i<j}^{3} \left(x_ix_j\displaystyle \sum_{p=1}^{3}t_{ip}t_{jp}+y_iy_j\displaystyle \sum_{p=1}^{3}t_{pi}t_{pj}\right)\right\}\right.\right.\nonumber\\
&&\left.\left.+\displaystyle\sum_{i=1}^2t_{2i}^2t_{3i}^2+\displaystyle \sum_{i\geqslant j}^2\displaystyle\sum_{p=j+1}^3t_{ij}^2t_{ip}^2+\frac{9}{4}\displaystyle\sum_{p=1}^3\displaystyle\sum_{i,j}^3\left\{(-1)^{\delta_{pi}}x_p+(-1)^{\delta_{pj}}y_p\right\}t_{ij}^2 \right\}\right],
\label{eq:ws_local_inseparable_w4}
\end{eqnarray}
where $\delta$ is the determinant of correlation matrix $T$ of the initial state $\rho_{12}$, $L_f=\frac{1}{6^6}( 3^6+ 2^6L_{\delta})$, $L_{\delta}=2\delta+3g_{33}+\frac{9}{4}L_5$, $g_3=(x_3+y_3)$, $L_1=\gamma_+-2(x^2_3+y^2_3)$, $L_2=\frac{9}{4}(t_{33}+\gamma_+-2x_3y_3)$, $L_3=x_3+y_3(\gamma_-+\frac{9}{4})$, $L_4=-L_2+\frac{3}{2}L_3$, $L_5=t_{33}L_1+L_4$, $g_{33}=(g_3+\frac{3}{2})C_{33}$, $S_q=(-1)^{1-\delta_{jq}}$, $\gamma_{\pm}=\|\vec{x}\|\pm\|\vec{y}\|$, $K_1=\gamma_-^2+\frac{9}{4}\gamma_+$, $K_2=3^8+6^4K_1+\frac{8}{9}\delta$ and $\delta_{ij}$ is the kronekar delta. Here $\|\vec{a}\|=\tr \left(a^{\dagger} a\right)$ with $\dagger$ denoting the Hermitian conjugate.
These nonlocal outputs $\tilde{\rho_{14}}$ and $\tilde{\rho_{23}}$ will be inseparable when,
\begin{equation}
W^l_3<0 \text{ or }  W^l_4 <0 \text{ and } W^l_2 \geqslant 0.
\label{eq:range_local_inseparability_appendix}
\end{equation}
\section*{Appendix-2: Inseparability range of desired outputs obtained using nonlocal cloners}
\noindent Here, we again evaluate the determinants $W_2$, $W_3$ and $W_4$ (as given in Eq.~\eqref{eq:w3-w4}) of the Peres-Horodecki criterion for the states $\tilde{\rho}_{12}$ and $\tilde{\rho}_{34}$ given by Eq.~\eqref{eq:gen_nonlocal_nonlocaloutputs1}, and denote them as $W^{nl}_2$, $W^{nl}_3$ and $W^{nl}_4$ respectively. The mathematical expressions of these determinants turn out to be the following,
\begin{eqnarray}
W^{nl}_2=\frac{1}{20^2}\left[5(5+6x_3)-9 \left(\displaystyle\sum_{i=1}^3\{t_{3i}^2+y_i\left(2t_{3i}+y_i\right)\}-x_3^2\right)\right],
\label{eq:ws_nonlocal_inseparable_w2}
\end{eqnarray}
\begin{eqnarray}
&& W^{nl}_3 = \frac{9}{20^3} \left[f_4+f_{33}\displaystyle \sum_{i=1}^{3}\left(t_{3i}+y_i\right)^2-\left\{\ell_-\displaystyle\sum_{i=1}^3\left( t_{i3}+ x_{i}\right)^2+\ell_+\displaystyle\sum_{i,j,k,l}^2S_+t_{ij}t_{kl}+3\displaystyle \sum_{i,j}^2[t_{ij}^2-(x_i-t_{i3})^2]\right.\right.\nonumber\\
&&\left.\left.-6\left(\displaystyle\sum_{i=1}^2\left\{(x_it_{i2}+y_it_{2i})+x_i(t_{1i}t_{3i}-t_{2i}t_{3i})+(-1)^{i+1}\displaystyle\sum_{j=1}^2[t_{ij}t_{j3}t_{3j}+x_iy_i(t_{ii}-t_{2i})]\right\}+\displaystyle\sum_{i\neq j}^2t_{ij}(t_{i3}+t_{3j})\right.\right.\right.\nonumber\\
&&\left.\left.\left.+\displaystyle\sum_{i,j}^2(-1)^{\delta_{ij}}y_1t_{ij}t_{j3}\right)\right\}\right],
\label{eq:ws_nonlocal_inseparable_w3}
\end{eqnarray}
\begin{eqnarray}
&& W^{nl}_4 =\frac{1}{20^4}\left[ f_5-18\displaystyle\sum_{i,j=1}^{3}\ell_{ij}t_{ij}^2+81\left(\displaystyle\sum_{i,j}^3\displaystyle\sum_{k,l}^3S_{\delta}t_{ij}^2t_{lk}^2+8\displaystyle\sum_{i,j}^2\displaystyle\sum_{k=i+1,l=j+1}^3t_{ij}t_{il}t_{kj}t_{kl}\right)+ 1080\displaystyle\sum_{i,j}^3t_{ij}x_iy_j\right.\nonumber\\
&&\left.  +324\left\{\displaystyle\sum_{l=2}^3\left(x_l^2\displaystyle\sum_{i}^3(t_{1i}^2-t_{li}^2)+y_l^2\displaystyle\sum_{i}^3(t_{i1}^2-t_{il}^2)\right)+2\left(\displaystyle\sum_{i,j}^3x_iy_jC_{ij} -\displaystyle\sum_{i}^2\displaystyle\sum_{j\neq i,k}^3(t_{ik}t_{jk}x_ix_j +t_{ki}t_{kj}y_iy_j)\right)\right\}\right],
\label{eq:ws_nonlocal_inseparable_w4}
\end{eqnarray}
where $f_{33}=(3(x_3+y_3+t_{33})-5)$, $f_3=\frac{1}{9}(5+3x_3)^2$, $S_+=(-1)^{i+j+k+l}$, $f_4=6y_3C_{33}-f_3f_{33}$, $f_5=-275-1080\delta$, $S_{\delta}=(-1)^{1-\max(\delta_{il},\delta_{jk})}$, $\ell_{\pm}=5\pm3t_{33}+3x_3$, $C_{ij}$ is the co-factor of $t_{ij}$ in correlation matrix $T$, and $\ell_{ij}$ are elements of coefficient matrix $[\ell_{ij}]=\left(\begin{smallmatrix} 43&25&25\\ 25&7&7\\7&7&7 \end{smallmatrix}\right)$. These desired output states $\tilde{\rho}_{12}$ and $\tilde{\rho}_{34}$ will be inseparable when,
\begin{equation}
W^{nl}_3<0 \text{ or }  W^{nl}_4 <0 \text{ and } W^{nl}_2 \geqslant 0.
\label{eq:range_nonlocal_inseparability_appendix}
\end{equation}
\end{widetext}

\end{document}